\documentclass[11pt]{article}
\pdfoutput=1
\usepackage{jheparxiv}

\usepackage[dvipsnames]{xcolor}
\usepackage{tikz-feynman}
\usepackage{amsmath}
\usepackage{amsfonts}
\usepackage{bm}
\usepackage{braket}
\usepackage{comment}
\usepackage[normalem]{ulem}
\usepackage{amssymb}   
\usepackage{dsfont}
\usepackage{mathtools}
\usepackage{soul}
\usepackage{subcaption}

\usepackage{hyperref}
\hypersetup{colorlinks= true, linkcolor=blue, citecolor=red}

\usetikzlibrary{decorations.pathmorphing,decorations.markings}
\tikzset{snake it/.style={decorate, decoration={snake,segment length=4}}}
\tikzset{smallsnake/.style={decorate, decoration={snake,segment length=2, amplitude=1}}}
\newcommand{\VP}{\braket{\mathcal{S}}} 
\newcommand{\Am}{\bar{\mathcal{M}}} 
\newcommand{\Mamp}{\mathcal{M}} 
\newcommand{\vertex}[1]{%
    \draw[black, thick] (#1,0)--(#1+1,0);
    \draw[black, thick] (#1+1,0)--(#1+2,0);
    \draw[snake it, thick] (#1+1,0)--(#1+1,-1);
}
\newcommand{\vertexL}[1]{%
    \draw[black, thick] (#1,0)--(#1+.75,0);
    \draw[snake it, thick] (#1+.75,0)--(#1+.75,-1);
}
\newcommand{\vertexR}[1]{%
    \draw[black, thick] (#1,0)--(#1+.75,0);
    \draw[snake it, thick] (#1,0)--(#1,-1);
}
\newcommand{\vertexM}[1]{%
    \draw[snake it, thick] (#1,0)--(#1,-1);
}
\newcommand{\RvertexM}[1]{%
    \draw[snake it, thick] (0,#1)--(-1, #1);
}
\newcommand{\propF}[1]{
        \draw[black, thick] (#1,0)--(#1+1,0);
}
\newcommand{\propR}[1]{
        \draw[line width=.5mm, red] (#1,0)--(#1+1,0);
}
\newcommand{\VpropF}[1]{
        \draw[black, thick] (0,#1)--(0,#1+1);
}
\newcommand{\VpropR}[1]{
        \draw[line width=.5mm,red] (0,#1)--(0,#1+1);
}

\newcommand{\dd}{\mathrm{d}}
\newcommand{\hdelta}{\hat{\delta}}
\newcommand{\hd}{\hat{\dd}}
\renewcommand{\[}{\begin{equation}\begin{aligned}}
\renewcommand{\]}{\end{aligned}\end{equation}}

\newcommand{\be}{\begin{equation}}
\newcommand{\ee}{\end{equation}}
\newcommand{\bi}{\begin{enumerate}}
\newcommand{\ei}{\end{enumerate}}
\newcommand{\ud}{{\mathrm{d}}}
\def\Put(#1,#2)#3{\leavevmode\makebox(0,0){\put(#1,#2){#3}}}
\renewcommand{\SS}{\mathcal{S}} 

\definecolor{allOrderBlue}{rgb}{0.4,0.5,1}
\DeclareMathOperator{\T}{T}
\newcommand{\alphaStarInv}{\alpha^{*-1}}

\newcommand{\rin}{{\rm in}}
\newcommand{\rout}{{\rm out}}
\newcommand{\bp}{{\bm p}}

\title{The ABCs of Amplitudes, Bogoliubov and Crossing \\
}
\author[1]{Rafael Aoude,}
\emailAdd{rafael.aoude@ed.ac.uk}
\author[1,2]{Asaad Elkhidir,}
\emailAdd{elkhidir@ihes.fr}
\author[1]{Anton Ilderton,}
\emailAdd{anton.ilderton@ed.ac.uk}
\author[1]{Donal O'Connell,}
\author[]{and}
\emailAdd{donal@ed.ac.uk}
\author[1]{Karthik Rajeev}
\emailAdd{karthik.rajeev@ed.ac.uk}

\affiliation[1]{Higgs Centre for Theoretical Physics,
School of Physics and Astronomy, \\
The University of Edinburgh, Edinburgh EH9 3JZ, Scotland, UK}
\affiliation[2]{Institut des Hautes Études Scientifiques, 91440 Bures-sur-Yvette, France}
\date{}
\abstract{
It is now common to describe classical backgrounds involving dynamical black holes with the production of gravitational radiation using the methods of scattering amplitudes.
In that light, we revisit the standard formulation of quantum field theory on a background.
We discuss the interpretation of Bogoliubov coefficients as generalised amplitudes, and explain how crossing, analyticity, and causality relate the relevant set of amplitudes.
When the background is itself a coherent state, we map these statements onto standard results in flat-space quantum field theory.
}

\begin{document}

\maketitle

\newpage

\section{Introduction}

The on-shell approach to scattering amplitudes has proven to be very fruitful over the last few decades. This programme led to a revolution~\cite{Bern:2022jnl} in precision for cross-sections at hadron colliders, and enabled the direct computation of multiloop properties of supergravity~\cite{Bern:2013uka,Bern:2018jmv,Adamo:2022dcm}.
Remarkably, emphasising the importance of on-shell simplifications led to deep new insight into the double copy~\cite{Bern:2008qj,Bern:2010ue,Bern:2010yg,Cachazo:2013hca,Monteiro:2014cda}, which is now widely established as a crucial aspect of gravity. 
Today, on-shell methods and the scattering amplitude programme are under intensive study in the context of classical gravitational-wave physics~\cite{Travaglini:2022uwo}.

A by-product of research on the on-shell approach to gravitational-wave physics has been a notable improvement in our understanding of the classical limit of scattering amplitudes~\cite{Kosower:2018adc,Cristofoli:2022phh,Bini:2024rsy,Elkhidir:2024izo,Biswas:2024ept,DeAngelis:2025vlf,Aoude:2024jxd}. A rich variety of classical phenomena are routinely computed using the methods of amplitudes; for example, the emission of classical radiation during scattering processes~\cite{Ilderton:2013tb,Herderschee:2023fxh,Brandhuber:2023hhy,Elkhidir:2023dco,Georgoudis:2023lgf}, taking into account the effects of black hole spin angular momentum~\cite{DeAngelis:2023lvf,Aoude:2023dui,Brandhuber:2024bnz,Bohnenblust:2023qmy,Bohnenblust:2025gir,Alaverdian:2025jtw}, and subtle effects associated with large gauge transformations~\cite{Georgoudis:2023eke}.

It is natural to ask how our newly-sharpened insight into the classical limit of quantum field theory (QFT) can be extended to semiclassical or quantum effects. This prompts us to revisit here the Bogoliubov approach to quantum field theory on a (classical) background. Formally, this approach involves writing the ladder operators for (say) the out Fock space as linear combinations of the ladder operators of the in Fock space. The expansion coefficients --- the Bogoliubov coefficients --- capture, e.g., the physics of pair-production due to time-dependent fields. But this is on-shell physics: the production of physical particles which propagate for macroscopic distances. It should then be possible to reformulate the story and properties of Bogoliubov coefficients entirely in terms of scattering amplitudes; this is the goal of our paper. (Some connections between amplitudes and Bogoliubov coefficients have already been explored in the literature~\cite{Aoki:2024bpj,Copinger:2024pai,Aoude:2024sve,Melville:2023kgd,Melville:2024ove,Ilderton:2025umd,Aoki:2025ihc}.)

There are many physical systems that motivate the study of quantum field theory in a background, including QCD in strong magnetic fields~\cite{Miransky:2015ava,Hattori:2016emy}, intense laser physics~\cite{Fedotov:2022ely}, magnetar environments~\cite{Turolla:2015mwa,Kaspi:2017fwg}, high-energy QCD and the color glass condensate~\cite{Iancu:2003xm}. The presence of background fields, in particular strong fields, can often lead to explicitly nonlinear and nonperturbative effects, including spontaneous particle production in both cosmological~\cite{Semren:2025dix} and terrestrial~\cite{Fedotov:2022ely} settings. QFT in a fixed background can also be regarded as the zeroth order term in the self force expansion (probe limit) of a fully dynamical system~\cite{Adamo:2023cfp,Cheung:2023lnj,Kosmopoulos:2023bwc}.

All of these topics can be approached from an on-shell perspective. Compared to the case of perturbative scattering in vacuum, however, we know little about general analytic properties of scattering amplitudes on backgrounds~\cite{Ilderton:2020rgk,Lippstreu:2025jit,Dunne:2025cyo}, where even low-point tree-level amplitudes can exhibit almost arbitrary functional complexity, nor about extensions of common on-shell methods (integration by parts identities, double copy, etc) beyond flat space~\cite{Adamo:2022dcm}. Motivated by this, and heavily influenced by discussions in the (relatively) recent scattering amplitudes literature~\cite{Caron-Huot:2010fvq,Mizera:2021fap,Mizera:2021ujs,Caron-Huot:2023ikn}, we will here study \emph{crossing} in background field theories.

We begin in Section~\ref{sec:Background} by explaining the overall setup. 
We consider a quantum field theory involving some sector whose dynamics generates large expectation values of some set of fields: this is the background. 
We assume that an additional field $\phi$ is coupled to this background sector, and we treat $\phi$ as a probe of the background sector.
We take the action to be quadratic in the probe,
and assume that the probe couples very weakly to all other quantum fields.
However, as the background expectations are large, the interaction of the probe with the background is non-negligible.
We discuss the basic structure of the amplitudes in this situation, and demonstrate that the Bogoliubov coefficients are indeed on-shell functions --- not quite amplitudes, but in-in generalisations of the usual in-out amplitudes.

The next two sections discuss the relationships between the Bogoliubov generalised amplitudes and the usual amplitudes in our setup.
We discuss crossing~\cite{Bros:1965kbd,Bros:1964iho,BROS_1986325} in Section~\ref{sec:crossing}, and in Section~\ref{sec:InOut_InIn} we connect in-out scattering amplitudes to the Bogoliubov in-in amplitudes by relating propagators with different causal boundary conditions.
In Section~\ref{sec:coherent} we discuss in more detail the case of backgrounds that admit a coherent state description, connecting crossing in background fields to crossing of vacuum amplitudes.  We conclude in Section~\ref{sec:Conclusion}.

\paragraph{Conventions}
We consistently absorb powers of $(2\pi)$ in measures and delta function as
\be
\hat{\dd}^n p  \coloneqq \frac{\dd^n p}{(2\pi)^n} \;,
\qquad
\hdelta^{n}(p)  \coloneqq (2\pi)^n \delta^n(p),
\ee
which combine in the following on-shell measure and on-shell delta function
\be
    \dd\Phi(p)  \coloneqq \frac{\hat{\dd}^3{\bm p}}{2E_{\bm p}} \;,
    \qquad
 \delta_\Phi(p-p')   \coloneqq\,2E_p\, \hdelta^{(3)}({\bm p} - {\bm p}') \;.
\ee
It will be useful for us to distinguish between background and flat space amplitudes. We will adopt the notation ${\cal M}$ for the former and ${\cal A}$ for the latter. We also use ${\cal S}$ and $S$ for the background and flat space S-matrix, respectively.

\section{Background QFT and Bogoliubov coefficients}
\label{sec:Background}
We begin with a broad overview of our setup. We are interested in a quantum field theory whose action has the general form
\[\label{eq:generalAction}
I = \int \dd^4 x \left( \mathcal{L}_f +  \partial_\mu \phi^\dagger  \partial^\mu \phi-  m^2 \phi^\dagger \phi + \mathcal{O}(x) \, \phi^\dagger(x)\phi(x) \right) \,,
\]
where $\mathcal{L}_f$ is independent of $\phi$ and describes some set of fields $f$.
The operator $\mathcal{O}(x)$ contains fields from $\mathcal{L}_f$ coupling (potentially via derivatives as in QED or gravity) locally and quadratically to the probe $\phi$. Now, we assume that the dynamics of $\mathcal{L}_f$ generates (perhaps via strong interactions, or via highly-populated initial states, or for other reasons) large expectation values $\Gamma(x) = \braket{f(x)}$ of the fields $f$: this is the ``background''\footnote{At this stage, there is no assumption that these backgrounds are ``classical'' in the sense of having negligible variance~\cite{Cristofoli:2021jas}.}. We then rewrite the fields as fluctuations about their expectation value, $f = \Gamma+\delta f$, and further assume that the coupling of the probe to the quantum fluctuation $\delta f$ is small, so that we may completely neglect it.  In effect this replaces all the quantum fields in $\mathcal{L}_f$ and $\mathcal{O}(x)$ by their expectation values.

A particularly simple example, which we call $J$ theory, is a purely scalar theory with one field $f(x)$, a real probe $\phi(x)$ and interaction
\[\label{eq:operator_expand}
\mathcal{O}(x) \, \phi^2(x) = g f(x)  \, \phi^2(x) = J(x) \, \phi^2(x) + g \, \delta f(x) \, \phi^2(x) \,,
\]
where $J(x)  \coloneq g \Gamma(x)$ is a large quantity, whereas $g$ itself is small so that we may drop the $g \, \delta f(x)$ term. This assumption, that $g$ is negligible but $J(x) = g \Gamma(x)$ is large, is critical in what follows; it means we avoid applying perturbation theory to the dynamics of the $f$ fields,  and also radiation of $f$ particles is suppressed by the small coupling\,\footnote{This approximation is routinely used in strong-field QED, where it is called the Furry expansion~\cite{Furry:1951bef}.}.  In this limit, $\phi(x)$ should be thought of as a probe, interacting with a fixed background $J(x)$; this allows us to access systems which may be difficult to treat using straightforward perturbation theory.
One caveat is that we assume that the only boundaries of our spacetime are those of Minkowski space.
In particular, applying our results to Hawking radiation requires some extra work to account for the additional boundary on the horizon.

With these approximations, the action for the probe particle in $J$ theory is
\[\label{eq:probeAction}
I_\phi = \frac{1}{2} \int \dd^4 x \left( \partial_\mu \phi \partial^\mu \phi-  m^2 \phi^2  + J(x)\phi^2(x) \right) \;.
\]
More generally, the coupling of $\phi$ with the background involves derivatives.

\subsection{Primitive processes}

As the action~\eqref{eq:generalAction} is Gaussian in $\phi(x)$ there are essentially three basic scattering processes in our theory.
These are the one-to-one scattering amplitude:
\[\label{eq:primitive1}
\Mamp(p \rightarrow k) &= \begin{tikzpicture}[scale=1, baseline={([yshift=-\the\dimexpr\fontdimen22\textfont2\relax] current bounding box.center)}]
\begin{feynman}
\vertex (vblob);
\vertex [above left=0.5 and 1 of vblob] (i1) {};
\vertex [above right=0.5 and 1 of vblob] (o1) {};
\diagram{
	(i1) -- [thick] (vblob) -- [thick] (o1);
};
\filldraw [color=white] (vblob) circle [radius=8pt];
\filldraw [fill=allOrderBlue] (vblob) circle [radius=8pt];
\end{feynman} \,,
\end{tikzpicture}
\]
and the pair-production amplitude and its conjugate, the pair-annihilation amplitude:
\[\label{eq:primitive2}
\Mamp(0 \rightarrow k_1 k_2) &= 
\begin{tikzpicture}[scale=1, baseline={([yshift=-\the\dimexpr\fontdimen22\textfont2\relax] current bounding box.center)}]
\begin{feynman}
\vertex (vblob);
\vertex [above right=0.5 and 1 of vblob] (i1) {};
\vertex [below right=0.5 and 1 of vblob] (o1) {};
\diagram{
	(i1) -- [thick] (vblob) -- [thick] (o1);
};
\filldraw [color=white] (vblob) circle [radius=8pt];
\filldraw [fill=allOrderBlue] (vblob) circle [radius=8pt];
\end{feynman}
\end{tikzpicture}\,,
\qquad
\qquad
\Mamp(k_1 k_2 \rightarrow 0) &= 
\begin{tikzpicture}[scale=1, baseline={([yshift=-\the\dimexpr\fontdimen22\textfont2\relax] current bounding box.center)}]
\begin{feynman}
\vertex (vblob);
\vertex [above left=0.5 and 1 of vblob] (i1) {};
\vertex [below left=0.5 and 1 of vblob] (o1) {};
\diagram{
	(i1) -- [thick] (vblob) -- [thick] (o1);
};
\filldraw [color=white] (vblob) circle [radius=8pt];
\filldraw [fill=allOrderBlue] (vblob) circle [radius=8pt];
\end{feynman}
\end{tikzpicture} \,.
\]
We can think of these processes as primitive processes, because more complicated amplitudes are products of these primitive building blocks: there are no other connected diagrams we can draw. So, for example, the $2 \rightarrow 6$ amplitude is built from products of $1\to1$, $2\to 0$ and $0\to 2$ amplitudes.  
We can therefore expect that the full $S$ matrix is itself built from these primitive processes.
To see that this is the case, we start with $J$ theory, with the action~\eqref{eq:probeAction}.
Although we focus on this example for concreteness, our arguments apply to the complete set of theories of equation~\eqref{eq:generalAction}, including the possibility of derivative couplings between the probe $\phi(x)$ and the fields $f(x)$.

In $J$ theory, the $S$ matrix is
\[
\SS 
&= \lim_{t \rightarrow \infty} U(t, -t) \,,
\]
where $U$ is the finite time (interaction picture) evolution operator\footnote{In the case of derivative coupling, we should either write the interaction Hamiltonian density in the exponent of equation~\eqref{eq:timeEvolutionJ}, or use covariant time-ordering. These are equivalent due to Matthews' theorem~\cite{Matthews:1949zz}, see also~\cite{Kaplanek:2025moq} for a helpful recent discussion.}
\[\label{eq:timeEvolutionJ}
U(t_1, t_0) = \T \exp \left[ i \int_{t_0}^{t_1} \dd t \int \dd^3 x \, J(x) : \phi^2(x) : \right] \,,
\]
in which we have explicitly indicated a normal-ordering of the $\phi$ operators.
It is evident that the $S$ matrix is, up to the time ordering, an exponential of a quadratic form in the ladder operators of the interaction picture quantum field $\phi(x)$.
To see that this persists once the time ordering is taken into account, consider writing the time evolution as a product of a large number $N$ of small time steps
\[\label{eq:evolStep}
U(t_1, t_0) = U(t_1, t_1 - \epsilon) U(t_1 -\epsilon, t_1 - 2 \epsilon) \cdots U(t_0 + \epsilon, t_0) \,,
\]
where $\epsilon = (t_1 - t_0) / N$ can be made very small by taking $N$ sufficiently large.
For small enough $\epsilon$ we may approximate
\[
U(t_i + \epsilon, t_i) \simeq \exp\left[ i \int_{t_i}^{t_i + \epsilon} \dd t \int \dd^3 x \, J(x) : \phi^2(x) : \right] \eqqcolon e^{A_i} \,.
\]
Then the product of two time evolution operators is
\[
U(t_i + \epsilon, t_i) U(t_j + \epsilon, t_j) = e^{A_i} e^{A_j} = e^{A_i + A_j + \frac12 [A_i, A_j] + \cdots}
\]
by the Baker–Campbell–Hausdorff (BCH) formula. 
As the $A_i$ are quadratic in the field $\phi$, the commutators $[A_i, A_j]$ are also quadratic in fields --- and hence in the creation and annihilation operators.
Similarly, all the nested commutators in the BCH formula are quadratic in $a(p)$ and $a^\dagger(p)$.
Multiplying up all the evolution operators in equation~\eqref{eq:evolStep} we conclude that the time evolution operator over a finite interval is itself quadratic in ladder operators.
Thus the $S$ matrix in $J$ theory is 
\[\label{eq:Sstructure}
\SS &= e^{i \theta} e^{i\hat{N}} \,,\\
\hat{N}&= \int \dd\Phi(p_1, p_2) \left( f_1(p_1, p_2) a^\dagger(p_1) a^\dagger(p_2) + f_2(p_1, p_2) a^\dagger(p_1) a(p_2) + f_1^*(p_1, p_2) a(p_1) a(p_2) \right) \,,
\]
where $\theta$ is a numerical 
phase. 
The same is true for all the theories of equation~\eqref{eq:generalAction} under our approximations; differences due to spacetime derivatives enter through the exact form of the functions $f_i$ in equation~\eqref{eq:Sstructure}. Complex fields require us to include creation and annihilation operators also for antiparticles, but otherwise the arguments are unchanged. 

Notice that the structure of the $S$ matrix in equation~\eqref{eq:Sstructure} is consistent with the three primitive processes: they are now captured by the functions $f_1(p_1, p_2)$, its conjugate, and $f_2(p_1, p_2) = f_2^*(p_2, p_1)$.
We see that indeed the physics of the theory involves disconnected scattering of individual particles from the past to the future, accompanied by creation and absorption of particle pairs.

\subsection{The Bogoliubov coefficients}

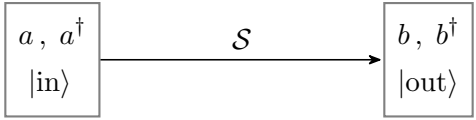
\begin{figure}[t]
\centering
\begin{tikzpicture}[auto]
\node (in) at (0,0) [rectangle,draw=black!50,thick,inner sep=5pt] {$\begin{aligned} &a \,,\; a^\dagger \\ &\,\,|\textrm{in} \rangle  \end{aligned}$};
\node (out) at (5,0) [rectangle,draw=black!50,thick,inner sep=5pt] {$\begin{aligned} &b \,,\; b^\dagger \\  &|\textrm{out} \rangle \end{aligned}$};
\draw [->,shorten >=1pt,>={Stealth[round]},semithick] (in.east) to node {$\SS$} (out.west);
\end{tikzpicture}
\caption{\label{fig:aSb} The $\SS$ matrix maps the in Fock space to the out Fock space.}
\end{figure}

The simple structure of the $S$ matrix~\eqref{eq:Sstructure} leads to a powerful relation between the ladder operators $a(p)$ and $a^\dagger(p)$ of the in Fock space and the corresponding annihilation $b(p)$ and creation $b^\dagger(p)$ operators of the out Fock space.
Both these sets of mode operators satisfy the usual canonical commutation relations
\[\label{mode-norm}
[a(p),a^\dagger(p')] &=  \delta_\Phi(p-p') \,, \\
[b(p),b^\dagger(p')] &= \delta_\Phi(p-p') \,.
\]
We define the vacuum of the in Fock space of (free) particles in the far past as
\begin{align}
&a(p) | \rin\rangle =  0 \;. 
\end{align}
Because time evolution sets up an isomorphism between the in and out spaces, as shown in figure~\ref{fig:aSb}, it is always the case that
\[\label{eq:b_is_SaS}
b(p) = \SS^\dagger a(p) \SS \,.
\]
It then follows that the $b(k)$ annihilate an `out vacuum' 
\[\label{eq:out_is_Sin}
\ket{\rout} = \SS^\dagger\ket{\rin},
\]
which is not, note, the time-evolved $\ket{\rin}$ vacuum.

In our case, there is an important simplification: we may use the explicit form~\eqref{eq:Sstructure} to calculate the out ladder operators in terms of the in ladder operators.
Indeed it is easy to see that the commutators $[a(p), \hat{N}]$ and $[a^\dagger(p), \hat{N}]$ are linear in $a$ and $a^\dagger$, so the Baker–Hausdorff lemma leads to
\[\label{eq:BogIntro}
b(p) = e^{-i\hat{N}} a(p) e^{i\hat{N}} = \int \dd \Phi(k) \left( \alpha(p,k)a(k) + \beta(p,k) a^\dagger(k)\right) \,.
\]
The quantities $\alpha(p,k)$ and $\beta(p,k)$ appearing here are the celebrated Bogoliubov coefficients.
They are clearly related to $f_1$ and $f_2$, but in a complicated way arising from the Baker-Campbell lemma.
Nevertheless, it is once again clear that they correspond in some way to the primitive processes~\eqref{eq:primitive1} and ~\eqref{eq:primitive2}.
In fact it is easy to extract $\alpha$ and $\beta$ from \eqref{eq:BogIntro} by taking overlaps as follows: 
\[
\label{eq:bogol-def}
&\alpha(p,k) \coloneqq \langle \rin| b(p) a^\dagger(k) | \rin \rangle =  \langle \rin| \SS^\dagger a(p) \SS a^\dagger(k) | \rin \rangle, \\
&\beta(p,k) \coloneqq \langle \rin| a(k) b(p) | \rin \rangle =  \langle \rin| a(k) \SS^\dagger a(p) \SS | \rin \rangle.
\]
We see that indeed there is a link to the on-shell primitive processes since $\alpha$ is an on-shell quantity involving the creation and subsequent annihilation of a particle, while $\beta$ involves the annihilation of two particles. 
However, the Bogoliubov coefficients involve \emph{two} insertions of the $S$ matrix, so they are not directly the primitive scattering amplitudes --- instead, they are generalised amplitudes~\cite{Schwarz:2019ggp,Schwarz:2019npn,Caron-Huot:2023vxl}; generalised in the sense that the boundary conditions are of in-in rather than in-out type.
We will explicitly see below that the propagators in the generalised amplitudes have a different $i\epsilon$ prescription to the usual Feynman $i \epsilon$.
In terms of our logic so far, the generalised Bogoliubov amplitudes differ from the more familiar amplitudes because it requires some work to rewrite the exponential of ladder operators in equation~\eqref{eq:Sstructure} into a normal ordered form from which we could identify the in-out amplitudes; it also requires work to connect the $f_i$ to $\alpha$ and $\beta$ explicitly through the Baker-Hausdorff lemma.
In the remainder of the paper, we will see in full detail how the Bogoliubov coefficients are related to scattering amplitudes --- and (by crossing) to themselves.

First, let us comment on another approach to the Bogoliubov coefficients.
In our class of theories, the equation of motion for the Heisenberg field of the probe is
\be
    \big(\partial^2 + m^2 \big)\phi_H(x) =  J(x) \phi_H(x) + J^\mu(x)\partial_\mu \phi_H(x) + J^{\mu\nu}(x)\partial_\mu\partial_\nu\phi_H(x) \;,
\ee
in which the $J$'s are some set of currents that encode all interactions with the external fields. This is a second-order partial differential equation, so we can expand the Heisenberg field in a set of positive and negative energy modes with boundary conditions adapted to either the asymptotic past ($P$) or future ($F$), in which the field becomes free. For our real scalar, this amounts to
\[
 \phi_H(x)&=\int \dd\Phi(k)\left[P(x,k)a(k)+P^*(x,k)a^{\dagger}(k)\right] \,,  \\
 &= \int \dd\Phi(k)\left[F(x,k)b(k)+F^*(x,k)b^{\dagger}(k)\right] \label{future}\,,
\]
where the mode functions obey
\[
    P(x,k) &\sim e^{-ik\cdot x} \;, \qquad  x^0\rightarrow -\infty\;, \\
    F(x,k) &\sim e^{-ik\cdot x} \;, \qquad x^{0}\rightarrow \infty\,.
\label{future2}\]
Therefore, the field operator behaves as 
\[\label{eq:phiinphiout}
\phi_H(x) \sim \phi_\textrm{in}(x) &= \int \dd \Phi(k) ( e^{-i k \cdot x} a(k) + e^{i k \cdot x} a^\dagger(k)) \;, \qquad x^0 \to - \infty \,,
\\
\phi_H(x) \sim \phi_\textrm{out}(x) =\SS^\dagger \phi_\textrm{in}(x) \SS &= \int \dd \Phi(k) ( e^{-i k \cdot x} b(k) + e^{i k \cdot x} b^\dagger(k)) \;, \qquad x^0 \to + \infty \,.
\]
We may hence extract the ladder operators from the in/out fields using, for example, 
\[
b(p) = i \int \dd^3 x\, e^{i p \cdot x} \frac{\overset{\leftrightarrow}{\partial}}{\partial x^0} \phi_\textrm{out}(x) \,.
\]
Inserting $\phi_\textrm{out}(x)$ in terms of the past form of the Heisenberg field, we immediately see that
\[
b(p) = {\lim_{x^0 \rightarrow  +\infty}} \int \dd\Phi(k) \left[ i \int\!\dd^3 x \, e^{i p \cdot x} \frac{\overset{\leftrightarrow}{\partial}}{\partial x^0} \left( P(x, k) a(k) + P^*(x, k) a^\dagger(k ) \right)\right] \,.
\]
This can be recognised as the Bogoliubov decomposition~\eqref{eq:BogIntro}:
\be
b(p) = \int \dd \Phi(k) \left( \alpha(p,k)a(k) + \beta(p,k) a^\dagger(k)\right)  \;,\label{eq:b_from_Bogola}
\ee
in which we have obtained expressions for the Bogoliubov coefficients in terms of the past modes:
\[\label{eq:crossingHint}
\alpha(p, k) &= 
{\lim_{x^0 \rightarrow  +\infty}}
\,i\!\int\! \dd^3 x \, e^{i p \cdot x} \frac{\overset{\leftrightarrow}{\partial}}{\partial x^0}  P(x, k), \\
\beta(p, k) &=   
{\lim_{x^0 \rightarrow  +\infty}}
\,i\!\int\!\dd^3 x \, e^{i p \cdot x} \frac{\overset{\leftrightarrow}{\partial}}{\partial x^0}  P^*(x, k) \,.
\]
We can similarly obtain
\be\label{eq:b_from_Bogolb}
a(p) = \int \dd \Phi(k) \left( \alpha^*(k,p)b(k) - \beta(k,p) b^\dagger(k)\right) \;,
\ee
and express the Bogoliubov coefficients in terms of the future modes, if we wish.
We will return to all these equations below.

Note that we will sometimes adopt a cleaner matrix notation, writing e.g.
\be
\begin{split}
b(p) &= \alpha(p, k) \cdot a(k) + \beta(p, k) \cdot a^\dagger(k) \,,
\end{split}
\ee
where the summation/integration over repeated momenta is implicit.
Note that the two Bogoliubov coefficients in \eqref{eq:crossingHint} are related by interchanging $P(x,k) \leftrightarrow P^*(x,k)$ in their respective expressions. This is our first hint for a crossing relationship between the two coefficients, which we explore in Section \ref{sec:crossing}. Before doing so, we discuss how these coefficients are related to scattering amplitudes in the remainder of this section.

\subsection{Amplitudes from Bogoliubov   coefficients}\label{subsec:amp_from_bogol}

The Bogoliubov relations~\eqref{eq:bogol-def} reveal surprising connections between amplitudes and generalised amplitudes. 
To see how this works, let us first be more specific about our definitions.
The primitive one-to-one amplitude, shown diagrammatically in equation~\eqref{eq:primitive1}, is
\[\label{eq:defOf1-1}
\Mamp(p \to k) \coloneqq& \braket{ \rin | a(k) \SS a^\dagger (p) | \rin} \\
=& \braket{ \rout | b(k) a^\dagger(p) | \rin } \,,
\]
using equations~\eqref{eq:b_is_SaS} and~\eqref{eq:out_is_Sin}. 
Unlike amplitudes in empty spacetimes, it is crucial to keep track of the difference between the incoming and outgoing vacua. Similarly, for the pair creation and annihilation amplitudes, we have
\[
\Mamp(0 \to k_1 k_2) \coloneqq& \braket{ \rout | b(k_1) b(k_2) | \rin} = \braket{ \rin | a(k_1) a(k_2) \SS | \rin} \,,\\
\Mamp(k_1 k_2 \to 0) \coloneqq& \braket{\rout | a^\dagger(k_1) a^\dagger(k_2) | \rin}  \,.
\]
Note that our definition of the `amplitude' $\mathcal{M}$ includes both forward scattering contributions as well as any momentum-conservation delta functions -- it is really the full $S$-matrix element.

To relate the one-to-one amplitude to a generalised amplitude, define the (matrix)
inverse $\alphaStarInv(k,p)$: 
\[
\int \dd\Phi(p) \, \alpha^*(q, p) \alphaStarInv(p,k) = \delta_\Phi(q-k) = \alpha^*(q,p) \cdot \alpha^{*-1}(p,k)\,.
\]
We emphasise that this inverse is not simply $1/\alpha^*(p,q)$. 
Now, with the help of the Bogoliubov relation~\eqref{eq:b_from_Bogolb} it is easy to see that
\[\label{eq:manipulatedBog}
\bra{\rout} b(k) = \int \dd\Phi(p) \bra{\rout} \, a(p) \alphaStarInv(p,k)  \,.
\]
Inserting this result into the definition~\eqref{eq:defOf1-1} of the one-to-one amplitude, it follows that
\[\label{eq:1-1amp_is_alpha_inverse}
\Mamp(p \to k) = \alphaStarInv(p,k) \VP  \,,
\]
where we have introduced the vacuum persistence amplitude
\[
\VP \coloneqq \langle \rin | \mathcal{S} | \rin \rangle = \langle \rout | \rin \rangle \,,
\]
which is non-trivial in the presence of background fields. The relation (\ref{eq:1-1amp_is_alpha_inverse}) is known --- but it is surprising viewing the left-hand-side as a standard in-out amplitude and the right-hand side as an in-in generalised amplitude (up to factors and the conjugate inverse, of course). We will discuss this relationship in detail in Section~\ref{sec:InOut_InIn}.

A very similar result holds for the pair-creation amplitude. Using equations~\eqref{eq:b_from_Bogola} and~\eqref{eq:manipulatedBog} one can verify that
\[\label{eq:pairCreationAinvB}
\Mamp(0 \to k_1 k_2) = \VP \, \beta(k_2, p) \cdot \alphaStarInv(p, k_1) \,.
\]
Again, this is a standard result, yet it is surprising from the point of view of amplitudes given different boundary conditions on the left and the right. 
Indeed, the Bose symmetry of the left-hand side is far from obvious on the right-hand side: this is a non-trivial relationship. Finally, we can also obtain the relation 
\[
\Mamp(k_1 k_2\rightarrow 0) = -\VP \,\alphaStarInv( k_1,p)\cdot \beta^*(p, k_2) \,.
\]

We see in  \eqref{eq:1-1amp_is_alpha_inverse} and \eqref{eq:pairCreationAinvB} that both the one-to-one and pair creation amplitudes are proportional to the vacuum persistence amplitude $\braket{\SS}$. It is often convenient to define `barred' amplitudes with this factor divided out, so $\Am := \Mamp /\VP$, upon which we have
\begin{align}
    \Am(p \to k)=\alphaStarInv(p,k)  \;,
    \qquad
    \Am(0\to k_1 k_2)=\beta(k_2, p) \cdot \alphaStarInv(p, k_1) \,.
\end{align}
Having seen how our (barred) amplitudes are related to the Bogoliubov coefficients, it is natural to ask how the inverse relations look. We will show shortly how they follow essentially from unitarity and the Gaussian nature of the system. Before doing so, we briefly review the topic of pair creation in our system, which illustrates the direct role of the Bogoliubov coefficients in observable quantities and emphasizes the Gaussian character of $\SS$.

\subsection{Vacuum pair creation}

A particularly interesting aspect of quantum field theory on a background is the phenomenon of pair creation from the vacuum.
This can be described in complete detail using the methods we have discussed so far by computing $\SS \ket{\rin}$. 
Diagrammatically, the relevant primitive amplitude is the pair creation amplitude $\Mamp(0 \to k_1 k_2)$, so it is natural to presume that
\[\label{eq:expectedFinalState}
\SS \ket{\rin} = \sum_{n=0}^\infty X_n \left( 
\begin{tikzpicture}[scale=1, baseline={([yshift=-\the\dimexpr\fontdimen22\textfont2\relax] current bounding box.center)}]
\begin{feynman}
\vertex (vblob);
\vertex [above right=0.5 and 1 of vblob] (i1) {};
\vertex [below right=0.5 and 1 of vblob] (o1) {};
\diagram{
	(i1) -- [thick] (vblob) -- [thick] (o1);
};
\filldraw [color=white] (vblob) circle [radius=8pt];
\filldraw [fill=allOrderBlue] (vblob) circle [radius=8pt];
\end{feynman}
\end{tikzpicture}
\right)^n \,,
\]
where $X_n$ are appropriate symmetry and normalisation factors. The following well-known calculation establishes that this is the case, and supplies the appropriate factors.

Beginning with the obvious equality $\SS a(p) \ket{\rin} = 0$, use equation~\eqref{eq:b_from_Bogolb} while eliminating the out mode operators in favour of $a$ and $a^\dagger$ using equation~\eqref{eq:b_is_SaS}. The result is
\[
\int \dd \Phi(k) \left( \alpha^*(k,p)a(k) - \beta(k,p) a^\dagger(k)\right) \SS \ket{\rin} = 0 \,.
\]
To extract an equation for the final state $\SS \ket{\rin}$, use the inverse $\alphaStarInv$:
\[
a(p) \SS \ket{\rin} = \int \dd\Phi(k_1) \, \Am(0 \to k_1, p) \, a^\dagger(k_1) \SS \ket{\rin} \,,
\]
where we exploited equation~\eqref{eq:pairCreationAinvB}.
It is then trivial to verify the solution, which is
\[\label{eq:squeezeFinalState}
\SS \ket{\rin} = \VP \exp \left(\frac{1}{2} \, a^\dagger(p_1) \cdot\Am(0 \rightarrow p_1,p_2)\cdot a^\dagger(p_2)\right) \ket{\rin} \,,
\]
and is perfectly consistent with equation~\eqref{eq:expectedFinalState}.
We see that the $S$-matrix simply acts as a squeezing operator on the vacuum state, with the pair-production amplitude $ \mathcal{M}(0 \rightarrow p_1,p_2)$ as a squeezing parameter
(see also~\cite{Copinger:2024pai}). 
The exact exponentiation of the pair-production amplitude is consistent with the (diagrammatically obvious) vanishing of $0 \rightarrow n$ amplitudes for odd $n$ and the factorisation of the same amplitude for even $n$.

An observable measure of pair production is the number of outgoing particles with momentum $k$
\[
\dd n(k) \coloneqq \braket{ \rin | b^\dagger(k) b(k) | \rin } \,.
\]
This could be computed using the squeezed state~\eqref{eq:squeezeFinalState}, but we choose a different route to illustrate a point which will be useful below.
Inserting a complete set of states $\ket{X}$ (which we take to be spanned by the incoming particles), we have
\[\label{eq:dn_def_2}
\dd n(k) = \sum_X \braket{ \rin | \SS^\dagger a^\dagger(k) \SS | X} \braket{ X| \SS^\dagger a(k) \SS | \rin} \,.
\]
Now, the overlap $\braket{ X| \SS^\dagger a(k) \SS | \rin}$ actually vanishes unless $\bra{X}$ is a single particle state. 
This is easily seen by using equation~\eqref{eq:b_from_Bogola}.
Indeed more generally
\[\label{eq:GaussianIdentity}
\langle \rin | a(p_1') \cdots a(p_n') \left( \SS^\dagger a(p) \SS \right) a^{\dagger}(p_1) \cdots a^{\dagger}(p_n) | \rin \rangle = 0, \quad \text{if} \,\,\, |n-n'| \neq 1.
\]
Returning to the number distribution of final particles, we now have
\[\label{eq:n_is_betasqr}
\dd n(k) &= \int \dd\Phi(p) \braket{ \rin | \SS^\dagger a^\dagger(k) \SS a^\dagger(p)| \rin} \braket{ \rin| a(p) \SS^\dagger a(k) \SS | \rin} \\
&= \int \dd\Phi(p) \, \beta^*(k, p) \beta(k, p) \,.
\]
Thus, Bogoliubov $\beta$ is directly and simply related to an observable number distribution. 

\subsection{Bogoliubov coefficients from amplitudes}\label{Sec:SnakeyWakey}
We now proceed to carry out the `inverse' of the calculations leading to \eqref{eq:1-1amp_is_alpha_inverse} and \eqref{eq:pairCreationAinvB}, deriving the Bogoliubov coefficients from amplitudes. We begin with expression~(\ref{eq:bogol-def}) for $\alpha$ and insert a complete set of states:
\begin{align}
    \alpha(p,q)&=\sum_{X}\langle \text{in}| \mathcal{S}^\dagger \ket{X}\bra{X} a(p) \mathcal{S} a^\dagger(q) | \text{in} \rangle\\\label{eq:alpha_statesum}
    &=\sum_{X}|\!\VP\!|^2\Am^*(0\rightarrow X)\Am(q\rightarrow p,X)\,,
\end{align}
where we again invoke the `barred' $\Am = \Mamp/\langle\SS\rangle$. 

It is clear that only states $X$ with an even number of particles contribute to the sums above. The amplitudes appearing are then of the form $\Am_{0\rightarrow 2n}$ and $\Am_{1\rightarrow 2n+1}$, both of which can in turn be written in terms of the basic amplitudes $\Am_{0\rightarrow 2}$ and $\Am_{1\rightarrow 1}$ in a Gaussian theory. Specifically, for bosonic fields, one has 
\begin{align}
    \Am(0\rightarrow q_1,q_2,\cdots,q_{2n})&=\sum_{\pi}\prod_{\forall \{j,k\}\in\pi} \Am(0\rightarrow q_{j},q_{k}),\\
    \Am(p\rightarrow q_1,q_2, \cdots,q_{2n+1})&=\sum_{i=1}^{2n+1}\Am(p\rightarrow q_i)\Am(0\rightarrow q_1,q_2,\cdots,\hat{q}_i,\cdots,q_{2n+1}),
\end{align}
where $\hat{q}_i$ indicates the operation of omitting $q_i$ and $\pi$ denotes a pairing partition (i.e., a partition of the set $\{1,2,\cdots,2n\}$ into $n$ disjoint subsets of size 2). The first equation above follows from expanding the exponential in \eqref{eq:squeezeFinalState}, while the second one follows in a similar way from the analogous equation for $\SS\ket{p}$.

To aid us in accounting for the different momentum contractions that arise from the state-sum in \eqref{eq:alpha_statesum}, we define the diagrams
\be
    \Am(p\rightarrow p')\equiv
    \raisebox{2pt}{
    \begin{tikzpicture}[baseline]
\draw[double] (-.3,0)--(.3,0);
\end{tikzpicture}}\quad ,
\qquad \qquad 
    \Am(0\rightarrow p,p')\equiv
    \raisebox{2pt}{\begin{tikzpicture}[baseline]
\draw[double] (0,.25) arc (90:270:.25);
\end{tikzpicture}}\quad\;.
\ee
With this, we will now show that $\alpha(p,p')$ may be written as a sum over connected diagrams, each of which is constructed from sewing together the above primitive building blocks, where by  sewing we mean 
\begin{align}
    \int \dd\Phi(q)\Am(p\rightarrow q)\Am^*(0\rightarrow p',q)= \begin{tikzpicture}[baseline]
\draw[double] (-.5,0)--(0,0);
\draw[double] (0,0) arc (90:-90:.25);
\draw[dashed] (0,.2) -- (0,-.3);
\end{tikzpicture}\;,\\
\int \dd\Phi(q)\Am(0\rightarrow p,q)\Am^*(0\rightarrow p',q)= \begin{tikzpicture}[baseline]
\draw[double] (0,.5) arc (90:-90:.25);
\draw[double] (0,0) arc (90:270:.25);
\draw[dashed] (0,.2) -- (0,-.25);
\end{tikzpicture}\;,
\end{align}
and so on. To illustrate how only connected diagrams enter the final expression for $\alpha$, consider the following subset of \textit{disconnected} diagrams that contribute to the right-hand side of \eqref{eq:alpha_statesum}:  
\begin{align}
\alpha\supset 
 |\VP|^2\left[  \begin{tikzpicture}[baseline]
\draw[double] (-.5,1)--(0,1);
\draw[double] (0,1) arc (90:-90:.25);
\draw[double] (0,-.5+1) arc (90:270:.25);
\draw[dashed] (0,.2+1) -- (0,-.75+1);
\end{tikzpicture}+
\begin{tikzpicture}[baseline]
\draw[double] (-.5,1)--(0,1);
\draw[double] (0,1) arc (90:-90:.25);
\draw[double] (0,-.5+1) arc (90:270:.25);
\draw[dashed] (0,.2+1) -- (0,-.75);
\draw[double] (0,-.3) circle (.15);
\end{tikzpicture} +
\frac{1}{2!}\left(\begin{tikzpicture}[baseline]
\draw[double] (-.5,1)--(0,1);
\draw[double] (0,1) arc (90:-90:.25);
\draw[double] (0,-.5+1) arc (90:270:.25);
\draw[dashed] (0,.2+1) -- (0,-1);
\draw[double] (0,-.3) circle (.15);
\draw[double] (0,-.75) circle (.15);
\end{tikzpicture}+
\begin{tikzpicture}[baseline]
\draw[double] (-.5,1)--(0,1);
\draw[double] (0,1) arc (90:-90:.25);
\draw[double] (0,-.5+1) arc (90:270:.25);
\draw[dashed] (0,.2+1) -- (0,-1);
\draw[double] (0,-.15) arc (90:-90:.3);
\draw[double] (0,-.75) arc (270:90:.1);
\draw[double] (0,-.55) arc (-90:90:.1);
\draw[double] (0,-.35) arc (270:90:.1);
\end{tikzpicture}\right)+...\right]=\begin{tikzpicture}[baseline]
\draw[double] (-.5,1)--(0,1);
\draw[double] (0,1) arc (90:-90:.25);
\draw[double] (0,-.5+1) arc (90:270:.25);
\draw[dashed] (0,.2+1) -- (0,-.75+1);
\end{tikzpicture},
\end{align}
where we have used the relation~\cite{Copinger:2024pai}
\begin{align}
  |\VP|^{-2}  = 1
  +\begin{tikzpicture}[baseline]
\draw[double] (0,0) circle (0.3) ;
\draw[dashed] (0,0.35)--(0,-0.35);
\end{tikzpicture}
+ \frac{1}{2!}\left[\begin{tikzpicture}[baseline]
\draw[double] (0,0) circle (0.5*3/5) ;
\draw[dashed] (0,0.6*3/5)--(0,-0.6*3/5);
\end{tikzpicture}\times \begin{tikzpicture}[baseline]
\draw[double] (0,0) circle (0.5*3/5) ;
\draw[dashed] (0,0.6*3/5)--(0,-0.6*3/5);
\end{tikzpicture}
+\begin{tikzpicture}[baseline]
\draw[double] (0,0.5*3/5) arc (90:270:0.25*3/5) ;
\draw[double] (0,0) arc (90:-90:0.25*3/5); 
\draw[double] (0,-.5*3/5) arc (90:270:0.25*3/5) ; 
\draw[double] (0,-1*3/5) arc (-90:90:0.75*3/5); 
\draw[dashed] (0,0.6*3/5)--(0,-1.2*3/5);
\end{tikzpicture}\right]+\cdots\,,
\end{align}
to cancel the factor of $|\langle \SS \rangle|^2$ against the sum of disconnected (cut) bubbles.
Repeating this procedure to others similar sets of \textit{disconnected} diagrams contributing to \eqref{eq:alpha_statesum}, where the factor $|\braket{{\rm in}|S|\rm in{}}|^2$ cancels against the bubbles, we finally arrive at
\begin{align}\label{eq:alpha_snakey}
\alpha=
\begin{tikzpicture}[baseline=-2.5]
\draw[double] (-.5,0)--(.5,0);
\end{tikzpicture}
+
 \begin{tikzpicture}[baseline=-2.5]
\draw[double] (-.5,0)--(0,0);
\draw[double] (0,0) arc (90:-90:.25);
\draw[double] (0,-.5) arc (90:270:.25);
\draw[dashed] (0,.2) -- (0,-.75);
\end{tikzpicture}
+
 \begin{tikzpicture}[baseline=-2.5]
\draw[double] (-.5,0)--(0,0);
\draw[double] (0,0) arc (90:-90:.25);
\draw[double] (0,-.5) arc (90:270:.25);
\draw[double] (0,-1) arc (90:-90:.25);
\draw[double] (0,-1.5) arc (90:270:.25);
\draw[dashed] (0,.2) -- (0,-1.75);
\end{tikzpicture}
+\,\,
\cdots .
\end{align}
This, of course, is a diagrammatic representation of the functional relation 
\begin{align}
    \alpha(p,p')= \Am(p,q)\cdot\left[\delta_{\Phi}(q-p')-\Am^*(0\rightarrow q,q')\cdot\Am(0\rightarrow p',q')\right]^{-1}\,,
\end{align}
where the inverse is to be understood as the matrix inverse of the expression inside the square brackets. The analogous expression for $\beta$ can be derived similarly, or more directly using the relation $\beta(p,k)=\Am(0\rightarrow p,q )\cdot\alpha^*(q,k)$, as follows from \eqref{eq:pairCreationAinvB} and \eqref{eq:alpha_snakey}. This yields
\begin{align}\label{eq:beta_snakey}
    \beta=
    \begin{tikzpicture}[baseline=-2.5]
\draw[double] (.5,0)--(0,0);
\draw[double] (0,0) arc (90:270:.25);
\draw[dashed] (0,.25)--(0,-.25);
\end{tikzpicture}
+
\begin{tikzpicture}[baseline=-2.5]
\draw[double] (.5,0)--(0,0);
\draw[double] (0,0) arc (90:270:.25);
\draw[double] (0,-.5) arc (90:-90:.25);
\draw[double] (0,-1) arc (90:270:.25);
\draw[dashed] (0,.25)--(0,-1.25);
\end{tikzpicture}
+
\begin{tikzpicture}[baseline=-2.5]
\draw[double] (.5,0)--(0,0);
\draw[double] (0,0) arc (90:270:.25);
\draw[double] (0,-.5) arc (90:-90:.25);
\draw[double] (0,-1) arc (90:270:.25);
\draw[double] (0,-1.5) arc (90:-90:.25);
\draw[double] (0,-2) arc (90:270:.25);
\draw[dashed] (0,.25)--(0,-2.25);
\end{tikzpicture}
+\,\,
\cdots ,
\end{align}
which is equivalent to
\begin{align}
    \beta(p,p')=\Am(0\rightarrow p,q_1)\cdot \Am^*(q_1\rightarrow p')\cdot\left[{\delta}_{\Phi}(q_2-q_1) -\Am^*(0\rightarrow q_1,q_3)\cdot\Am(0\rightarrow q_2,q_3)\right]^{-1}\,.
\end{align}
It is worth emphasizing what has been demonstrated here. By recognizing that Bogoliubov coefficients are inclusive observables, we have shown that one can fully and exactly account for the contributions of all possible states into which the vacuum and a single-particle state may evolve within the fixed-background approximation. In that respect, these relations are special cases of formally exact functional equations connecting amplitudes to in-in objects in background QFT. 
We remark that functional relations such as these offer a way to systematically go beyond the fixed-background approximation, as was used in~\cite{Copinger:2024pai} to derive leading-order backreaction effects in strong-field QED. 

Let us return to the average number of particles that we calculated in \eqref{eq:n_is_betasqr}. In the spirit of the preceding discussion, one could also think of $\dd n(k)$ as an inclusive observable. Specifically
\begin{align}\label{eq:dn_def_3}
\dd n(k) &= \sum_X \braket{ \rin | \SS^\dagger| X} \braket{ X|a^\dagger(k)a(k) \SS | \rin} \\
&= \sum_n n\times \mathcal{M}^*(0\rightarrow 2n)\mathcal{M}(0\rightarrow 2n) ,
\end{align}
where the sum in the second line is schematic. As shown in \cite{Copinger:2024pai}, this series can be evaluated explicitly to yield
\begin{align}\label{eq:dn_def_4}
\int\dd n(k) &= \Am(0\rightarrow k,q_1)\Am^*(0\rightarrow q_1,q_2)\left[1-\Am(0\rightarrow q_2,q_3)\Am^*(0\rightarrow q_3,k)\right]^{-1}\,.
\end{align}
While the above expression is written explicitly in terms of the amplitudes, it is considerably more involved, requiring functional inverses and multiple integrals. In contrast, \eqref{eq:n_is_betasqr} provides a much more \emph{direct} evaluation, highlighting the significance of the Bogoliubov coefficients in the background QFT context. 

In light of this, we now turn to the analyticity properties underlying these coefficients. In particular, the next section will show that $\beta$ is uniquely determined by the Bogoliubov $\alpha$ coefficient, which is conceptually simpler: $\alpha$ arises from a (generalised) one-to-one process, and in fact, as shown in equation~\eqref{eq:1-1amp_is_alpha_inverse}, all the information is in the one-to-one amplitude.

\section{Crossing}\label{sec:crossing}

Current research on scattering amplitudes naturally emphasises the physical importance of the amplitudes themselves, often viewing them as the primary objects one wishes to compute.
However, we have already seen in equations~\eqref{eq:1-1amp_is_alpha_inverse} and~\eqref{eq:pairCreationAinvB} that the Bogoliubov coefficients determine the (primitive) amplitudes in our context.
Therefore, it is enough to determine $\alpha$ and $\beta$.
In this section, we will show that crossing relates these two coefficients --- so in fact it is sufficient to determine only one of them.
We achieve this by considering various field correlators and using LSZ reduction to relate them to on-shell (generalised) amplitudes.

\subsection{Response functions and generalised amplitudes}\label{sec:response}
Following~\cite{Caron-Huot:2010fvq,Caron-Huot:2023ikn}, but now in the context of quantum field theory in backgrounds, we consider retarded products of field operators,
\[
\mathcal{R}\{ \phi(x_2) \phi(x_1) \}  := \theta(x_2^0 - x_1^0 )[\phi(x_2) , \phi(x_1)]  \;,
\]
and the `response function' 
\[\label{eq:MomentumSPaceResponseFunction}
R(p_1,p_2) \equiv \int \dd^4 x_1 \dd^4 x_2 \, i e^{ i p_2 \cdot x_2} (\partial^2_{x_2}+m^2) \, ie^{i p_1 \cdot x_1} (\partial^2_{x_1}+m^2)\bra{ \text{in}} \mathcal{R} \{ \phi_H(x_2), \phi_H(x_1)\} \ket{\text{in}} \;,
\]
where the momenta $p_1^\mu $ and $p_2^\mu$ are not necessarily real or on-shell.
It is clear, though, that when the momenta are taken real and on-shell, the combination of the Fourier transform in $R(p_1,p_2)$ and Klein-Gordon operator in (\ref{eq:MomentumSPaceResponseFunction}) amounts to LSZ-reduction of a retarded, rather than time-ordered, correlator. 

Let us identify the physical content of the response function $R(p_1,p_2)$ for the case of on-shell momenta. We take $p_1^2=p_2^2=m^2$ with $p_1^0 >0$ and $p_2^0 >0$, i.e.~both momenta become those of physical particles, and the exponents are those associated with outgoing particles in LSZ reduction. 
To perform the integrals over $x_1^\mu$ in~\eqref{eq:MomentumSPaceResponseFunction}, we recall that 
\[
\int \dd^4 x_1 \, i e^{i p_1 \cdot x_1 } (\partial^2_{x_1}+m^2) f(x_1) = \int \dd^4 x_1 \frac{\partial}{\partial x_1^0}  \left( i e^{i p_1 \cdot x_1 } \frac{\overset{\leftrightarrow}{\partial}}{\partial x_1^0} f(x_1)  \right) \,,
\]
for any (smooth) function $f(x_1)$ which has the property that its spatial support is finite\footnote{As usual in LSZ, we may include wavepackets for the scattering states to ensure this spatial localisation.}.
Consequently, the $x_1^0$ integral in equation~\eqref{eq:MomentumSPaceResponseFunction} is exact, generating the boundary terms
\[\label{x1-integral-2}
    \int \dd^3 {\bf x}_1 \, e^{i p_1 \cdot x_1} \frac{\overset{\leftrightarrow}{\partial}}{\partial x_1^0} \, \theta(x_2^0 - x_1^0) \braket{ \rin | [\phi_H(x_2), \phi_H(x_1)] | \rin} \bigg|^{x_1^0 = \infty }_{x_1^0 = -\infty} \;.
\]
The boundary at large $x_1^0$ vanishes on account of the theta function leaving
\be\begin{split}
 - \int \dd^3 {\bf x}_1 \, e^{i p_1 \cdot x_1}\bra{ \text{in}} [\phi_H(x_2)\frac{\overset{\leftrightarrow}{\partial}}{\partial x_1^0} \phi_H(x_1)] \ket{\text{in}} \bigg|_{x_1^0 = -\infty}  &= i \bra{\text{in}}[\phi_H(x_2),a(p_1)]\ket{\text{in}}  \\
&= - i\bra{\text{in}} a(p_1) \phi_H(x_2)\ket{\text{in}}\;,
\end{split}
\ee
We proceed to the integrals over $x^\mu_2$, which amount to a standard LSZ reduction, generating boundary terms at $x_2^0=\pm\infty$, of which only that in the future contributes,
\be\begin{split}\label{x1-integral-3}
    R(p_1,p_2) &= i \int\!\ud^4x_2\, e^{i p_2\cdot x_2} (\partial_{x_2}^2+m^2)  \bra{\text{in}} a(p_1) \phi_H(x_2)\ket{\text{in}} \\
    &= \bra{\text{in}}a(p_1)\big(b(p_2)-a(p_2)\big)\ket{\text{in}} \\
    & = \bra{\text{in}}a(p_1) b(p_2) \ket{\text{in}} = \beta(p_2,p_1) \;,
\end{split}
\ee
and we thus find, using the definition (\ref{eq:bogol-def}), that $R(p_1,p_2)$ 
reduces at on-shell momenta to Bogoliubov $\beta$.

Now, suppose we instead evaluate $R$ at the same $p_1$, but $p_2 \to - p_2$, so that $p_2$ is now the on-shell momentum of an incoming particle. The calculation of the $x_1$ integrals in $R$ proceeds precisely as before, thus we can jump to the analogue of (\ref{x1-integral-3}):
\be\begin{split}\label{x1-integral-4}
    R(p_1,-p_2) &= i \int\!\ud^4x_2\, e^{-i p_2\cdot x_2} (\partial^2_{x_2}+m^2)  \bra{\text{in}} a(p_1) \phi_H(x_2)\ket{\text{in}} \\
    &= \bra{\text{in}}a(p_1)\big(a^\dagger(p_2)-b^\dagger(p_2)\big)\ket{\text{in}} \\
    & = \delta_\Phi(p_1-p_2) - \alpha^*(p_2,p_1) \;,
\end{split}
\ee
again using~(\ref{eq:bogol-def}). Thus, the response function also encodes Bogoliubov $\alpha$. Note that the delta function term removes the trivial forward-scattering contribution from (minus)~$\alpha^*$.

\subsection{Relating $\alpha$ and $\beta$}\label{Sec:GeneralCrossing}
We have found that $\beta(p_2,p_1)$ in (\ref{x1-integral-3}) and $\alpha(p_2,p_1)$ (minus the forward scattering contribution) in (\ref{x1-integral-4}) are obtained by evaluating the \emph{same} response function $R$ for different arguments.
This is in complete analogy with the crossing relationship between vacuum (generalised) amplitudes discussed in~\cite{Caron-Huot:2023ikn} (see their Eq.~2.14).
It therefore seems appropriate to view the relationship between $\alpha$ and $\beta$ as a form of crossing.

Crossing relations between amplitudes (and generalised amplitudes~\cite{Caron-Huot:2023vxl,Caron-Huot:2023ikn}) are not proven in general, although the validity of crossing in four-dimensional vacuum amplitudes is a common belief in the community. 
The difficulty with any proof lies in demonstrating that the various integrals converge along appropriate paths in a complexified space of on-shell kinematic invariants, which could relate (generalised) amplitudes with crossed amplitudes.
We will shed no light on convergence in this paper: we assume that the relevant integrals converge in an appropriate domain\footnote{In a completely general background, the integrals may not be analytic in the energy.
However, the backgrounds of interest to us arise from the dynamics of a unitary quantum field theory. We then expect that our amplitudes and response functions inherit analyticity properties from the full theory, up to singularities associated with the cut-off of our Gaussian effective theory. See eg.~\cite{Mizera:2021fap,Mizera:2023tfe,Correia:2025enx} for related discussions.}.
What we will discuss is that the $p \leftrightarrow -p$ relation between the Bogoliubov coefficients $\alpha(p,k)$ and $\beta^*(p,k)$ apparent in equation~\eqref{eq:crossingHint} is a form of crossing.

In more detail, we begin by observing that since $R(p_1,p_2)$ is the Fourier transform of a retarded correlator, it is analytic in the upper half plane $p_2^0$ plane by causality.  Suppose, then, that beginning with $R(p_1,p_2)$ with both momenta on-shell, we deform the energy $p_2^0$ (positive) through the upper half-plane to $-p_2^0$ (negative). Since ${\bm p}_2$ remains fixed, the response function is then analytically continued from $R(p_1,p_2)$ to $R(p_1,-\tilde{p}_2)$, in which ${p}^\mu_2 = (p_2^0, {\bm p}_2)$ and $\tilde{p}^\mu_2 = (p_2^0, -{\bm p}_2)$, the parity-reversed momentum. 
This analytic continuation is an off-shell version of crossing: in vacuum amplitudes, such continuations are a key ingredient in establishing that an \emph{on-shell} analytic continuation also exists, as we will explore in Section~\ref{sec:Standard_Crossing} (see also Appendix A of~\cite{Caron-Huot:2023ikn} for a detailed discussion).
Thus, subject to our assumptions, we have analytically continued Bogoliubov $\beta(p_2,p_1)$ to Bogoliubov $\alpha^*(\tilde{p}_2,p_1)$, which we write as the crossing equation 
\[\label{eq:BogoliubovCrossingEquation}
\Big[ \beta(p_2,p_1) \Big]_{\curvearrowleft} = \delta_\Phi(p_1 - \tilde{p}_2) - \alpha^*({\tilde p}_2,p_1) \,.
\]
We could similarly have considered $R(-p_1,p_2)$ from the outset ($p_1^2=m^2$ and $p_1^0>0$), and tracked the behaviour of the response function in different regions of $p_2^0$, but this does not give us any new information. One finds explicitly
\be\begin{split}
    R(-p_1,p_2) &= \alpha(p_2,p_1) - \delta_\Phi(p_2-p_1) \;, \\
    R(-p_1,-p_2) &= - \beta^*(p_2,p_1) \;,
\end{split}
\ee
consistent with simply taking the complex conjugate of (\ref{x1-integral-3}) and (\ref{x1-integral-4}), and the resulting crossing relation is just the conjugate of (\ref{eq:BogoliubovCrossingEquation}).

We therefore find that crossing between the two Bogoliubov coefficients follows from analyticity properties of retarded products. Note that the analytic continuation of $\beta$ does not reproduce the trivial forward-scattering part of $\alpha^*$; this is consistent with (i) the precise form of LSZ being used in the response function, which removes disconnected diagrams, and (ii) the fact that $\beta$ cannot contain a forward-scattering contribution.

The prescription to deform the momentum $p_1^0$ while leaving $p^\mu_2$ untouched corresponds to crossing a \emph{single} leg between the past and future. The background `assists' the corresponding physical processes by providing any appropriate missing energy-momentum, but the required amount can clearly differ between different scattering processes, suggesting that the background also undergoes some implicit crossing. In Section~\ref{sec:coherent} we will clarify this, connect our discussion more firmly to the crossing of vacuum amplitudes, and make the crossing role of the background explicit, by considering backgrounds which may be represented as coherent states. 
Here we discuss the analogue of the crossing equation (\ref{eq:BogoliubovCrossingEquation}) for ordinary scattering amplitudes.

\subsection{Crossing in amplitudes: entire functions}

It is natural, from an amplitude perspective, to consider time-ordered rather than retarded products,
\[
\mathcal{T}\{ \phi(x_2) \phi(x_1) \}  := \theta(x_2^0 - x_1^0 )\phi(x_2) \phi(x_1) +  \theta(x_1^0 - x_2^0 )\phi(x_1) \phi(x_2) \;.
\]
The natural analogue of the response function (\ref{eq:MomentumSPaceResponseFunction}) is the LSZ-reduction of a time-ordered \emph{in-out} correlator, rather than in-in,
\[\label{eq:TOrderedAmplitude}
T(p_1,p_2) \equiv \int \dd^4 x_1 \dd^4 x_2 \, i e^{ i p_2 \cdot x_2} (\partial^2_{x_2}+m^2) \, ie^{i p_1 \cdot x_1} (\partial^2_{x_1}+m^2)\bra{ \text{out}} \mathcal{T} \{ \phi_H(x_2), \phi_H(x_1)\} \ket{\text{in}} \;,
\]
which must return an ordinary scattering amplitude when evaluated for on-shell momenta. Let us then proceed to evaluate $T(p_1,p_2)$ much as we did before. We fix $p_1$ to be on-shell (with $p_1^0>0$), and perform the $x_1^\mu$ integrals, arriving at
\be\label{eq:T-partially-reduced}
    T(p_1,p_2) = i \int\!\ud^4 x_2 \, e^{ip_2\cdot x_2}\bra{\text{out}}b(p_1) (\partial^2_{x_2}+m^2)\phi_H(x_2)\ket{\text{in}} \;.
\ee
This is analogous to the expression (\ref{x1-integral-3}) for the response function $R(p_1,p_2)$. We again evaluate $T$ for two choices of momenta. Taking $p_2$ on-shell with positive energy, we have
\be\begin{split}\label{eq:T_is_M_0to2}
    T(p_1,p_2) &= 
    i \bra{\text{out}}b(p_1) \big(-i b(p_2) + i a(p_2)\big) \ket{\text{in}} \\
    &= \bra{\text{out}}b(p_1)b(p_2)\ket{\text{in}} \\
    & = \bra{\text{in}}a(p_1)a(p_2) \mathcal{S}\ket{\text{in}} = \mathcal{M}(0\to p_1 p_2)\;,
\end{split}
\ee
the pair production amplitude. Similarly, taking $p_2\to -p_2$, we obtain
\be\label{eq:T_is_M_1to1}
     T(p_1,-p_2) =\mathcal{M}(p_2 \to p_1) - \delta_\Phi(p_2-p_1)\langle\mathcal{S}\rangle \;,
\ee
the one-to-one scattering amplitude, with a forward-scattering contribution subtracted.

However,  because $T$ is time-ordered rather than retarded, even in vacuum we can no longer straightforwardly invoke analyticity to infer crossing relations between the one-to-one and pair production amplitudes. 
To make progress, we can consider the special case in which the external backgrounds $\Gamma$ are compactly supported, and in particular turn on at some finite initial time $t_i$ and turn off\,\footnote{For gauge fields and metrics, we assume there is a gauge choice such that the potentials or metric perturbations also vanish.} at some later time $t_f$. This makes it clear (as schematically indicated in Fig.~\ref{fig:aSb}) that we can set up well-defined scattering problems between `in' and `out' regions, with associated `in' and `out' Hilbert spaces, where the field $\phi_H(x)$ is free.

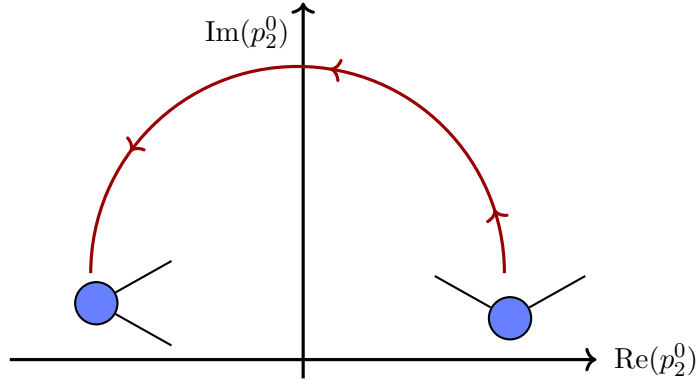
\begin{figure}
\centering
\def\xr{3.5}
\def\yr{3}
\begin{tikzpicture}[thick]
  \draw[very thick,->] (-\xr-0.4,0) -- (\xr+0.4,0) node [left=-1.5] {$\text{Re}(p_2^0)$};
  \draw[very thick,->] (0,-\yr+2.75) -- (0,\yr+1.75) node[below left=+0.05] {$\text{Im}(p_2^0)$};
  
  \begin{feynman}
    \vertex (vblob) at (\xr-0.75,\yr-2.45);
    \vertex [above left=0.5 and 1 of vblob] (i1) {};
    \vertex [above right=0.5 and 1 of vblob] (o1) {};
    \diagram{(i1) -- [thick] (vblob) -- [thick] (o1);};
    \filldraw [color=white] (vblob) circle [radius=8pt];
    \filldraw [fill=allOrderBlue] (vblob) circle [radius=8pt];
   \end{feynman}

    \begin{feynman}
      \vertex (vblob) at (-\xr+0.75,\yr-2.25);
      \vertex [above right=0.5 and 1 of vblob] (i1) {};
      \vertex [below right=0.5 and 1 of vblob] (o1) {};
      \diagram{
        (i1) -- [thick] (vblob) -- [thick] (o1);
      };
      \filldraw[color=white] (vblob) circle [radius=8pt];
      \filldraw[fill=allOrderBlue] (vblob) circle [radius=8pt];
    \end{feynman}

    \draw[xshift=5, very thick,red!60!black,decoration={markings,mark=between positions 0.1 and 1 step 0.35 with \arrow{>}},postaction={decorate}]  (\xr-1.0,\yr-1.85) arc (0:180:\yr-0.25);
   
\end{tikzpicture}
     \caption{
     \label{fig:entire}
     Crossing from one-to-one to zero-to-two amplitudes in any background-field theory with suitably compact fields, as outlined in~\eqref{eq:EntireCrossingEquation}.
     }
\end{figure}

The equations of motion for $\phi$, which follow from (\ref{eq:generalAction}), then imply that the correlator entering the definition of $T(p_1,p_2)$ in (\ref{eq:T-partially-reduced}), also vanishes in the in- and out-regions. It follows that the time integral in (\ref{eq:T-partially-reduced}) is convergent for \emph{any} choice of $p_2^0$, and so  $T(p_1,p_2)$ is an \emph{entire} function of $p_2^0$, hence analytic in $p_2^0$ (assuming, as before, convergence at spatial infinity).
We can therefore consider any complex deformation of $p_2^0$ from positive to negative values, analytically continuing $T(p_1,p_2)$ to $T(p_1,-\tilde{p}_2)$ (with ${\tilde p}_2^\mu = (p_2^0,-{\bm p}_2)$ the parity-reversed momentum, as before). We then have the crossing relation
\[\label{eq:EntireCrossingEquation}
\Big[ \mathcal{M}(0\to p_1 p_2) \Big]_{\curvearrowleft} = \mathcal{M}({\tilde p}_2 \to p_1) - \delta_\Phi({\tilde p}_2-p_1)\langle \mathcal{S}\rangle \;,
\]
between the pair production and one-to-one amplitudes (with, as for the generalised amplitudes previously considered, the subtraction of a forward-scattering term).
This is illustrated in Fig.~\ref{fig:entire}.
Diagrammatically, the two amplitudes are related by taking one leg and switching it from the past to the future, changing the direction of the spatial momentum -- \textit{this is crossing}. Hence, in any background-field theory with suitably compact fields, crossing relations between amplitudes are a direct result of analyticity. 

In fact, the crossing relation above can be derived from \eqref{eq:BogoliubovCrossingEquation}, assuming convergence of the momentum integrals appearing in products of Bogoliubov coefficients (such as $\beta \cdot \alpha^{*-1}$). With this assumption, the time-ordered crossing relation applies to a broader class of backgrounds than those of compact support. To see this, we first rewrite the left-hand side of \eqref{eq:EntireCrossingEquation} as 
\begin{align}
   \Big[\mathcal{M}(0\to p_1 p_2) \Big]_{\curvearrowleft}=  \Big[\beta(p_1,p) \cdot \alpha^{*-1}(p,p_2) \Big]_{\curvearrowleft}\langle \mathcal{S}\rangle\,.
\end{align}
We can now use \eqref{eq:BogoliubovCrossingEquation}, to cross $p_2$ in the above, to arrive at
\begin{align}
    \Big[\beta(p_2,p) \cdot \alpha^{*-1}(p,p_1) \Big]_{\curvearrowleft}= \Big[\delta_\Phi({\tilde p}_2-p_1)-\alpha^*(\tilde{p}_2,p)\Big] \cdot \alpha^{*-1}(p,p_1)\,.
\end{align}
Expanding the right-hand side, and using $\mathcal{M}(p_1\rightarrow p_2)=\alpha^{*-1}(p_1,p_2)\langle \mathcal{S}\rangle$ we immediately recover \eqref{eq:EntireCrossingEquation}.

The preceding discussions highlight the role of retarded versus time-ordered boundary conditions in LSZ reduction, and of the distinct vacua $\ket{\text{in}}$ and $\ket{\rout}$ in our theory. Bogoliubov coefficients and scattering amplitudes descend from, respectively, retarded in-in correlators and time-ordered in-out correlators\footnote{We could additionally consider reduction of eg.~in-in \emph{time-ordered} products or in-out \emph{retarded} products. These will, however, reduce to combinations of the Bogoliubov coefficients and scattering amplitudes above.}. These are, though, related by identities such as
\[\label{eq:RetardedTimeOrderedIdentity}
\mathcal{R}\{ \phi(x_1) \phi(x_2) \}  = \mathcal{T}\{ \phi(x_1) \phi(x_2)\} - 
 \phi(x_2)%
 \phi(x_1)  \;,%
\]
and their higher-point generalizations: applying LSZ to the LHS of \eqref{eq:RetardedTimeOrderedIdentity} yields Bogoliubov coefficients, while the same truncation applied to the RHS yields (cuts of) amplitudes.
There are thus many relations between  Bogoliubov coefficients and amplitudes, as we have seen in  Section~\ref{subsec:amp_from_bogol} and Section~\ref{Sec:SnakeyWakey}.

We now turn to a more detailed investigation of how the different boundary conditions that give rise to Bogoliubov coefficients and scattering amplitudes correspond, in the perturbative approach, to a precise difference in the choice of propagators.

\section{Causality: retarded versus Feynman}
\label{sec:InOut_InIn}

In this section, we illustrate how Bogoliubov coefficients arise from solutions of the field equations satisfying retarded boundary conditions, whereas scattering amplitudes are obtained from solutions obeying Feynman boundary conditions, using $J$ theory as an example.

\subsection{Bogoliubov coefficients and amplitudes in $J$ theory}\label{sec:J_theory}

We return to $J$ theory, with action (\ref{eq:probeAction}),  describing a massive scalar $\phi(x)$ coupled to a scalar background $J(x)=g\,\Gamma(x)$. We start by explicitly constructing, in this theory, the response function~\eqref{eq:MomentumSPaceResponseFunction}. For $k^\mu$ on-shell ($k^0>0$) and $q^\mu$ arbitrary, we consider the  response function $R(-k,q)$ in the form (\ref{x1-integral-3}), i.e.~
\be
\label{eq:MomentumSPaceResponseFunction-reprise2}
    R(-k,q) = i\int\! \dd^4 x \, e^{i q \cdot x}  (\partial^2+m^2)\langle \text{in}| \phi(x) a^\dagger(k)\ket{\text{in}} \;.
\ee
(For notational convenience, we have swapped from indexed momenta to momenta $k$ and $q$.)

Now, as written in (\ref{eq:MomentumSPaceResponseFunction-reprise2}), the correlation function (i) clearly describes a one-particle state, (ii) obeys the equations of motion since $\phi(x)$ does, and (iii) asymptotes to $e^{-ik\cdot x}$ in the past. It is thus equal to the $P(x,k)$ mode of \eqref{future2}, satisfying the equation of motion of $J$ theory, 
\[\label{eq:JtheoryEOM}
(\partial^2 +m^2) \phi(x) = g \Gamma(x) \phi(x) \;,
\]
with incoming boundary condition $P(x,k) \sim e^{ -i k\cdot x}$ as  $x^0 \rightarrow - \infty$. The response function is thus simply the LSZ reduction of $P(x,k)$ with momentum $q_\mu$. As already established in Section~\ref{sec:response} and Section~\ref{Sec:GeneralCrossing}, though, taking $q \to \pm p$ on-shell, this amputation recovers the Bogoliubov coefficients, i.e.
\be
\label{eq:reduce-phi-k}
\begin{split}
    R(-k,p) &= i\int\! \dd^4 x \, e^{i p \cdot x}  (\partial^2+m^2)\, P(x,k) = \alpha(p,k)-\delta_\Phi(k-p) \;, \\
     R(-k,-p) &= i\int\! \dd^4 x \, e^{-i p \cdot x}  (\partial^2+m^2)\, P(x,k) = -\beta^*(p,k) \;.
\end{split}
\ee
Using (\ref{eq:b_from_Bogola})--(\ref{eq:b_from_Bogolb}) this may equivalently be written as the statement that $P(x,k)$ evolves to
\[\label{eq:LSonshell}
P(x,k) \sim \int \dd \Phi(p) \left(e^{-i p \cdot x} \alpha(p,k) + e^{i p \cdot x} \beta^*(p,k) \right),
\]
in the far future -- this will be useful below.

To compute the Bogliubov coefficients perturbatively in the coupling $g$, we start with the following standard Lippmann-Schwinger approach to solving $P(x,k)$:
\[
P(x,k) = e^{ -i k\cdot x} + \sum_n P^{(n)}(x,k) \;,
\]
in which $P^{(n)}(x,k)$ is the $\mathcal{O}(g^n)$ term in the perturbative series.
The retarded boundary conditions on $P(x,k)$, recall~\eqref{future2}, are imposed by setting $ P^{(n)}(x,k) \rightarrow 0$ as $ x^0 \rightarrow  - \infty$. 
Solving the equations of motion quickly leads to the following result:
\[\label{eq:LSsolution}
P^{(n)}(x,k) =  \int \hd^4 p \, e^{- i p \cdot x} \, (-g)^n\int  \hd^4 p_n\cdots \hd^4 p_1\, \hat{\delta}^4(p_n-p)\prod_{i=1}^{n}\frac{\tilde{\Gamma}(p_{i}-p_{i-1})}{(p^2_{i}-m^2)_{R}}\, \;,
\]
where $p_{i=0}\equiv k$ and $(p^2-m^2)_{R}\equiv (p^0 + i \epsilon)^2 - \bp^2-m^2$ enforces the retarded boundary conditions. The LSZ reduction of $P(x,k)$ immediately leads to the following expression for the response function
\begin{align}\label{eq:J_theory_R_n}
  R(-k,p)|_{\mathcal{O}(g^n)}\equiv R^{(n)}(-k,p)= - i(-g)^n \int  \hd^4 p_n\cdots \hd^4 p_1\, \hat{\delta}^4(p_n-p)\frac{\prod_{i=1}^{n}\tilde{\Gamma}(p_{i}-p_{i-1})}{\prod_{i\neq n}(p^2_{i}-m^2)_{R}}  \,.
\end{align}

The Bogoliubov coefficients follow directly, in view of (\ref{eq:LSonshell}), as the overlaps of $P^{(n)}(x,k)$ with $e^{\pm i p\cdot x}$. The resulting expression takes the form:
\[\label{eq:J_theory_bogol_n}
&\alpha(p,k)|_{\mathcal{O}(g^n)}\equiv \alpha^{(n)}(p,k) = - i(-g)^n \int  \hd^4 p_n\cdots \hd^4 p_1\, \hat{\delta}^4(p_n-p)\frac{\prod_{i=1}^{n}\tilde{\Gamma}(p_{i}-p_{i-1})}{\prod_{i\neq n}(p^2_{i}-m^2)_{R}}, \\
&\beta^*(p,k)|_{\mathcal{O}(g^n)} \equiv \beta^{*(n)}(p,k) =  i(-g)^n \int  \hd^4 p_n\cdots \hd^4 p_1\, \hat{\delta}^4(p_n+p)\frac{\prod_{i=1}^{n}\tilde{\Gamma}(p_{i}-p_{i-1})}{\prod_{i\neq n}(p^2_{i}-m^2)_{R}}.
\]
The Bogoliubov coefficients are, therefore, seen to be directly related to the response functions $R(-k,\pm p)$, for $p$ on-shell, as dictated by the LSZ prescription in \eqref{eq:reduce-phi-k}.

We can similarly show that the time-ordered \emph{in-out}, and barred, correlator ${\bar T} := T /\VP$ takes the form
\begin{align}\label{eq:J_theory_F_n}
 {\bar{T}^{(n)}(k,-p)=} - i(-g)^n \int  \hd^4 p_n\cdots \hd^4 p_1\, \hat{\delta}^4(p_n-p)\frac{\prod_{i=1}^{n}\tilde{\Gamma}(p_{i}-p_{i-1})}{\prod_{i\neq n}(p^2_{i}-m^2)_{F}}  \,,
\end{align}
where $(p^2-m^2)_{F}$ denotes Feynman boundary conditions. From this, expressions for the amplitudes can be readily obtained using \eqref{eq:T_is_M_0to2} and \eqref{eq:T_is_M_1to1}. The solution to the equation of motion in this context is often referred to as the `Feynman solution'. In fact, the corresponding amplitudes can be extracted directly from the all-orders Feynman solution, as illustrated in~\cite{Ilderton:2025umd} in the case of Hawking radiation.

It is perhaps useful to give a more diagrammatic explanation of some of the preceding results. Continuing with $J$ theory, we have the simple Feynman rules
\begin{align}
    \begin{tikzpicture}[baseline=-.75cm]
        \vertex{-1}
        \node at (-1.2,0) {$p$};\node at (1.2,0) {$p'$};
    \end{tikzpicture}\quad
    &= \quad ig\tilde{\Gamma}(p'-p)\,,\\
    \begin{tikzpicture}
        \propF{-1}
        \node at (-0.5,.5) {$p$};
    \end{tikzpicture}\quad
    = \quad iG_F(p)\qquad&;\qquad
    \begin{tikzpicture}
        \propR{-1}
        \node at (-0.5,.5) {$p$};
    \end{tikzpicture}\quad
    = \quad i G_R(p) \; ,
\end{align}
where the retarded and Feynman propagators are
\[\label{eq:Retarded&FeynmanPropagators}
G_R(p) = \frac{1}{(p^2 -m^2)_R}, \quad  G_F(p) = \frac{1}{(p^2 -m^2)_F} \;.
\]
It is now simple to give the diagrammatic representation of the Bogoliubov coefficients as:
\begin{align}\label{eq:bogDiagrams}
  \alpha^{(n)}(p',p)=
  \begin{tikzpicture}[baseline]
        \vertexL{-1}
        \propR{-.25}
        \vertexM{.75}
        \draw[line width=.5mm,dashed,red] (.75,0)--(1.5,0);
        \vertexM{1.5}
        \propR{1.5}
        \vertexR{2.5}
        \node at (-1,.4) {$p$};
        \node at (.25,.4) {$p_1$};
        \node at (2,.4) {$p_{n-1}$};
        \node at (3,.4) {$p'$};
    \end{tikzpicture} \;,
    \quad
    -\beta^{*(n)}(p',p)=
\begin{tikzpicture}[baseline=.25cm]  
\draw[black, thick] (0,-1)--(.75,-1);
\RvertexM{-1}
\VpropR{-1}
\RvertexM{-1+1}
\draw[line width=.5mm,dashed, red] (0,0)--(0,.75);
\RvertexM{.75}
\VpropR{.75}
\RvertexM{.75+1}
\draw[black, thick](0,1.75)--(.75,1.75);
\node at (1,2) {$p'$};\node at (1,-.75) {$p$};
\node at (-.75, -.5) {$p_1$}; \node at (-.75, 1.25) {$p_{n-1}$};
\end{tikzpicture}\,.
\end{align}
The analogous decomposition for amplitudes is obtained by replacing $G_R$ with $G_F$:
\begin{align}\label{eq:ampDiagrams}
  \Am^{(n)}(p\rightarrow p')=
  \begin{tikzpicture}[baseline]
        \vertexL{-1}
        \propF{-.25}
        \vertexM{.75}
        \draw[thick,dashed] (.75,0)--(1.5,0);
        \vertexM{1.5}
        \propF{1.5}
        \vertexR{2.5}
        \node at (-1,.4) {$p$};
        \node at (.25,.4) {$p_1$};
        \node at (2,.4) {$p_{n-1}$};
        \node at (3,.4) {$p'$};
    \end{tikzpicture} \;,
    \quad
    \Am^{(n)}(0\rightarrow p',p)=
\begin{tikzpicture}[baseline=.25cm]  
\draw[black, thick] (0,-1)--(.75,-1);
\RvertexM{-1}
\VpropF{-1}
\RvertexM{-1+1}
\draw[thick, dashed] (0,0)--(0,.75);
\RvertexM{.75}
\VpropF{.75}
\RvertexM{.75+1}
\draw[black, thick](0,1.75)--(.75,1.75);
\node at (1,2) {$p'$};\node at (1,-.75) {$p$};
\node at (-.75, -.5) {$p_1$}; \node at (-.75, 1.25) {$p_{n-1}$};
\end{tikzpicture} \;.
\end{align}
Notice the close correspondence between the generalised amplitudes (Bogoliubov coefficients) in equation~\eqref{eq:bogDiagrams} and the amplitudes in equation~\eqref{eq:ampDiagrams}: the difference is simply the boundary condition ($i\epsilon$ prescription) defining the propagators.

\subsection{A simple example}
To explicitly understand the discussion above, it is useful to have an exact result against which to compare. We therefore restrict to the `impulsive' background
\begin{align}
    \Gamma(x)=\gamma_0\delta(x^0)\Rightarrow \tilde{\Gamma}(q)=\gamma_0\hat{\delta}({\bm q})
\end{align}
where $\gamma_0$ is a constant. The past and future modes defined in (\ref{future})--(\ref{future2}) can be found exactly in this case:
%
\begin{align}
\label{P-F-impulsive}
    P(x,p) &=e^{-ip\cdot x}\theta(-x^0) + \bigg[
        e^{-ip\cdot x}\left(1+\frac{ig \gamma_0}{2E(\bp)}\right)-\frac{ig \gamma_0}{2E(\bp)}e^{-i\bp\cdot x}e^{iE(\bp)x^0}\bigg]\theta(x^0) \\
    F^*(x,p)& =
        e^{ip\cdot x}\theta(x^0) + \bigg[
        e^{ip\cdot x}\left(1-\frac{ig \gamma_0}{2E(\bp)}\right)+\frac{ig \gamma_0}{2E(\bp)}e^{i\bp\cdot x}e^{-iE(\bp)x^0}\bigg]\theta(-x^0) \;,
\end{align}
%
where $E(\bp)=\sqrt{\bp^2+m^2}$. With this, the overlap integrals defining the Bogoliubov coefficients can also be computed,
\begin{align}
\alpha(p',p) =\hat{\delta}_{\Phi}(p-p')\left(1+\frac{i g \gamma_0}{2E(\bp)}\right) \;,
\qquad
\beta^*(p',p) = \hat{\delta}_{\Phi}(p+p')\left(-\frac{ig\gamma_0}{2E(\bp)}\right) \;.
\end{align}
Notice that these results are $\mathcal{O}(g)$--exact, implying that the perturbative series described previously truncates at this order. To see this, we start with the leading order correction in the expansion of $\alpha$, which reduces to
\begin{align}
    \alpha^{(1)}(p',p)&=ig \gamma_0\hat{\delta}(\bp'-\bp)=\hat{\delta}_{\Phi}(p'-p)\left(\frac{ig \gamma_0}{2E(\bp)}\right)\,.
\end{align}
At next-to-leading order, $\alpha^{(2)}(p',p)$ evaluates to
\begin{align}
\label{pic-R}
    \begin{tikzpicture}[baseline]
        \vertexL{-1}
        \propR{-.25}
        \vertexR{.75}
        \node at (-1,.5) {$p$};
        \node at (.25,.5) {$p_1$};
        \node at (1.5,.5) {$p'$};
    \end{tikzpicture}= ig^2\gamma_0^2\hat{\delta}(\bp-\bp')\int_{-\infty}^{\infty}\! \frac{\ud (p_1)_0}{2\pi}\frac{1}{((p_1)_{0}+i\epsilon)^2-E^2(\mathbf{p})}=0
\end{align}
where we have used the delta functions in the vertices to perform the integral over spatial components of $p_1$. The remaining integral vanishes, since the residues at the two poles cancel each other. By iterating this argument, all higher-order terms $\alpha^{(n)}(p',p)$ with $n\geq 2$ likewise vanish, consistent with the exact result derived from the mode functions (\ref{P-F-impulsive}). A similar calculation confirms the truncation of Bogoliubov $\beta^*$.

A notable feature of this example is that the amplitudes, in contrast to the Bogoliubov coefficients, receive non-zero contributions at all orders in the coupling $g$. The leading-order calculation coincides with that of the Bogoliubov coefficients, so we turn immediately to next-to-leading order for the one-to-one amplitude, $\Am^{(2)}(p\rightarrow p')$. In analogy to (\ref{pic-R}) we have
\begin{align}
    \begin{tikzpicture}[baseline]
        \vertexL{-1}
        \propF{-.25}
        \vertexR{.75}
        \node at (-1,.5) {$p$};
        \node at (.25,.5) {$p_1$};
        \node at (1.5,.5) {$p'$};
    \end{tikzpicture}
    &= ig^2\gamma_0^2\hat{\delta}(\bp-\bp')\int_{-\infty}^{\infty}\!\frac{\ud(p_1)_0}{(2\pi)}\frac{1}{(p_1)_{0}^2-E^2(\bp)+i\epsilon}\\
    &=\hat{\delta}_{\Phi}(p'-p)\left(\frac{ig \gamma_0}{2E(\bp)}\right)^2\,,
\end{align}
where, in contrast to the calculation of $\alpha^{(1)}$, using the Feynman propagator ensures that the $(p_1)_0$ integral receives a contribution from only the single pole in the lower half-plane, yielding a non-zero final answer. At higher orders, the delta functions at the vertices ensure that all spatial integrations in $\Am^{(n)}(p\rightarrow p')$ collapse into a single momentum-conserving delta function, leaving only the time components to integrate. Each remaining integral contributes one residue per propagator, resulting in
\begin{align}
    \Am^{(n)}(p\rightarrow p')=\hat{\delta}_{\Phi}(p'-p)\times\left(\frac{ig \gamma_0}{2E(\bp)}\right)^n\,.
\end{align}
The perturbative series for $\Am(p\rightarrow p')$ is therefore geometric, and can be resummed to
\begin{align}
    \Am(p\rightarrow p')&=\hat{\delta}_{\Phi}(p'-p)\left(1-\frac{ig \gamma_0}{2E(\bp)}\right)^{-1}\,.
\end{align}
This result also illustrates  \eqref{eq:1-1amp_is_alpha_inverse}. The calculation of the pair creation amplitude follows similarly, leading to
\begin{align}
    \Am(0\rightarrow p,p')&=\hat{\delta}_{\Phi}(\tilde{p}+p')\left(\frac{ig \gamma_0}{2E(\bp)}\right)\left(1-\frac{ig \gamma_0}{2E(\bp)}\right)^{-1}\,,
\end{align}
which recovers the $\beta\cdot \alpha^{*-1}$ form of the amplitude. 

Of course, once we know that the Bogoliubov coefficients terminate at $\mathcal{O}(g)$, we expect the amplitudes to receive contributions from all orders, since they can be written as $\beta\cdot \alpha^{*-1}$. What the perturbative computation has shown is how this structure emerges: 
the pole in each Feynman propagator contributes a single residue, and the resulting chain of residues stacks itself into exactly the geometric series predicted by the non-perturbative expressions. This neatly illustrates how the choice of propagator determines the perturbative expansion and how it matches onto the series structure of the corresponding on-shell quantities -- the Bogoliubov coefficients on the one hand, and the amplitudes on the other.

\section{Coherent State Backgrounds}\label{sec:coherent}

In this section, we restrict to the specific class of background which represent (linearised) solutions of vacuum wave equations, and can thus equivalently be described as asymptotic coherent states of the appropriate field (photons, gluons, gravitons, etc.)~\cite{Kibble:1965zza,Frantz:1965, Brown:1968dzy}. Coherent states have become a central tool in understanding the physics of radiation following~\cite{Glauber:1951zz,sudarshan:1963es,Glauber:1963tx}, and see~\cite{Endlich:2016jgc,Ilderton:2017xbj,Cristofoli:2021vyo,Aoude:2023fdm,Adamo:2025vzv} for recent amplitudes-focussed applications. 
For us, coherent states offer an opportunity to connect our results to crossing in vacuum amplitudes.

We take a step back and consider our scalar field $\phi(x)$ now coupled to a \emph{dynamical}, massless, spin-$s$ field $f$ (where we naturally have in mind photons, gravitons, etc). The field operator has the mode expansion 
\[\label{eq:GammaModeExpansion}
f^{\{ \mu_1, \cdots,\,\mu_s\}}(x) =\int \dd \Phi(k) \left(  \varepsilon^{\{ \mu_1, \cdots,\,\mu_s\}}_\eta(k) a_\eta(k) e^{ - i k \cdot x}  + 
\varepsilon^{\{ \mu_1, \cdots,\,\mu_s\} *}_\eta(k)
a^\dagger_\eta (k)e^{i k \cdot x} \right) 
\]
where $\{\mu_1, \cdots,\,\mu_s \}$ denotes a (possibly empty) set of Lorentz indices and $\eta$ denotes the helicity (implicitly summed over when repeated).  
Note that we write the annihilation operator of this field as $a_\eta(k)$.

We will consider scattering of number states of $\phi$, as usual and as above, together with \emph{coherent} states of the massless field $\ket{\gamma}$ defined by~\cite{Glauber:1951zz,Glauber:1963tx}
\[\label{eq:CoherentStateDef}
| \gamma \rangle = \mathbb{D}[\gamma] | 0 \rangle \equiv \exp \bigg[ \int \dd \Phi(k) ( \gamma_\eta(k) a^\dagger_\eta(k)-\gamma_\eta^*(k)a_
\eta(k))\bigg] | 0 \rangle, 
\]
in which  we call $\gamma_\eta(x)$ the waveshape and $\mathbb{D}[\gamma]$ a displacement operator, obeying
\be
\label{displace}
\mathbb{D}^\dagger [\gamma] \, a_\eta(k) \, \mathbb{D}[\gamma] = a_\eta(k) + \gamma_\eta(k) \;.
\ee
As such, coherent states generate non-zero expectation values of the field:  
\[
\label{eq:CoherentBackground}
\begin{split}
\Gamma^{\{ \mu_1, \cdots,\,\mu_s\}}(x) &
\coloneqq
\langle \gamma| f^{\{ \mu_1, \cdots,\,\mu_s\}}(x) | \gamma \rangle \\
&= \int \dd \Phi(k) \left(  \varepsilon^{\{ \mu_1, \cdots,\,\mu_s\}}_\eta(k) \gamma_\eta(k) e^{ - i k \cdot x}  + 
\varepsilon^{\{ \mu_1, \cdots,\,\mu_s\}}_\eta(k) \gamma^*_\eta (k)
e^{i k \cdot x} \right) \;,
\end{split}
\]
and it can be verified that $\Gamma^{\{ \mu_1, \cdots,\,\mu_s\}}(x)$ is a classical solution of the appropriate linearised vacuum equations $\Box \, \Gamma^{\{ \mu_1, \cdots,\,\mu_s\}} = 0$. 

Using (\ref{displace}), we can recast scattering with a coherent state $\ket{\gamma}$ in terms of a slightly modified $S$ matrix.
As a first step, we write a generic amplitude involving the coherent state as
\[
\braket{\gamma\, p'_1 \cdots | S | \gamma \, p_1 \cdots} = \braket{p'_1 \cdots |\mathbb{D}^\dagger [\gamma] S \mathbb{D}[\gamma] | p_1 \cdots}
\,, \]
where the $p_i$ and $p'_i$ are states of $\phi$.
%
%
Observe, using equation~\eqref{displace}, that this corresponds to the background-field approach at the level of the action with precisely the background of equation~\eqref{eq:CoherentBackground}~\cite{Frantz:1965,Kibble:1965zza}.
We further simplify in the spirit of Section~\ref{sec:Background} (see \eqref{eq:operator_expand}), taking the coupling $g$ between $\phi$ and $f$ fields to be small, while assuming that $J = g\Gamma$ is nevertheless large enough to warrant a non-perturbative treatment. Consequently, contributions from internal lines and loops of $f$ particles are suppressed by powers of $g$; in the language of \eqref{eq:operator_expand}, these correspond to the effects of the $g\, \delta f \,\phi^2$ term. Restricting to leading order in this scheme, $\mathbb{D}^\dagger [\gamma] \, S \, \mathbb{D}[\gamma]$ reduces precisely to the $S$-matrix for the $\phi$ field in the fixed background $\Gamma$; in short
\[
\mathbb{D}^\dagger [\gamma] \, S \, \mathbb{D}[\gamma] \simeq \SS 
\]
in agreement with the definition adopted in Section~\ref{sec:Background} and Section~\ref{sec:crossing}. We can then define Bogoliubov coefficients by
\[\label{eq:CoherentBogoliubov}
&\alpha(p,k)  
= \langle \gamma| S^\dagger a(p) S a^\dagger(k) | \gamma \rangle
\simeq \langle \text{in}| \SS^\dagger a(p) \SS a^\dagger(k) | \text{in} \rangle \;, \\
&\beta(p,k) =  \langle \gamma| a(k) S^\dagger a(p) S | \gamma \rangle 
\simeq
\langle \text{in}| a(k) \SS^\dagger a(p) \SS | \text{in} \rangle \;,
\]
which match (\ref{eq:bogol-def}) in the strict background-field limit.

\subsection{Coherent state crossing}\label{sec:coherent_state_X}

The expressions in \eqref{eq:CoherentBogoliubov} offer a concrete connection between background-field and flat-space amplitudes. This allows us to explore how the background-field crossing discussed in Section~\ref{sec:crossing} is related to the more familiar crossing of scattering amplitudes in flat space. As before, our discussion revolves around the response function 
\[\label{eq:CoherentStateResponseFunction}
   R(-k,q) = i\int\! \dd^4 x \, e^{i q \cdot x}  (\partial^2+m^2)\bra{\gamma}\phi(x) a^\dagger(k)\ket{\gamma} \;.
\]
where we have used the coherent state $\ket{\gamma}$ to represent the background field. We now focus on the case of scalar coherent states for simplicity, although the discussion readily generalises to higher spin fields. Under the assumptions stated above --- wherein contributions from internal lines and loops of the $f$ particles are suppressed--- the correlator $\bra{\gamma}\phi(x) a^\dagger(k)\ket{\gamma}$ solves \eqref{eq:JtheoryEOM}, the equation of motion of the $J$ theory, with the source now given by 
\begin{align}
    \Gamma(x)=\bra{\gamma}f(x)  \ket{\gamma}\,.
\end{align}
We can further simplify matters by defining $\gamma(-p) \equiv \gamma^*(p)$ for $p^0<0$, allowing us to write
\[\label{eq:CoherentStateJ}
\bra{\gamma}f(x)  \ket{\gamma} = \int \hd^4 p \,e^{- i p \cdot x} \,\hdelta(p^2) \gamma(p)\,,
\]
 from which we readily identify 
\[
\tilde{\Gamma}(p) = \hdelta(p^2) \gamma(p)\,,
\]
where $\tilde{\Gamma}(-p)=\tilde{\Gamma}^*(p)$. This means that we can immediately import the results in Section~\ref{sec:J_theory} to study the response function in the coherent state background.

The key distinction in the present case is that the background $\tilde{\Gamma}(p)$ is composed of on-shell massless particles. An immediate consequence is that the $\mathcal{O}(g)$ contribution to the Bogoliubov coefficients vanishes, since it corresponds to an on-shell three-point after LSZ reduction.
At higher orders, the on-shell source is responsible for generating non-trivial symmetry factors in the perturbative expressions of the Bogoliubov coefficients. Therefore, we consider the leading non-trivial contribution to the response function, at $\mathcal{O}(g^2)$. From \eqref{eq:J_theory_R_n}, we find
\begin{align}
    R^{(2)}(-k,p)=- i(-g)^2 \int  \hd^4 q_2\,\hd^4 q_1\, \hat{\delta}^4(k+q_1+q_2-p)\frac{\tilde{\Gamma}(q_2)\tilde{\Gamma}(q_1)}{[(k+q_1)^2-m^2]_{R}}\,,  
\end{align}
where we have changed the integration variable to $q_1=p_1-k$ and $q_2=p_2-p_1$.

We now rewrite the integral in a form that is manifestly symmetric under the exchange $q_1 \leftrightarrow q_2$, thereby expressing the response function in the following suggestive form:
\begin{align}\label{eq:R_g2_scalar}
    R^{(2)}(-k,p)= \frac{1}{2}\int  \hd^4 q_2\,\hd^4 q_1\, \hat{\delta}^4(k+q_1+q_2-p)\,\tilde{\Gamma}(q_2)\tilde{\Gamma}(q_1)\times i\tilde{\mathcal{A}}(k , q_1 , q_2 ,p)\,,  
\end{align}
where
\[\label{eq:OffShellAmplitude}
\tilde{\mathcal{A}}(k , q_1 , q_2 ,p) \equiv (ig)^2\left( \frac{1}{(q_1+k)^2-m^2} + \frac{1}{(q_2+k)^2-m^2}\right).
\]
Observe that the quantity $\tilde{\mathcal{A}}$ is a vacuum single off-shell current as the momentum $p$ is off-shell and the energies $q_1^0$ and $q_2^0$ can take positive or negative values. The tilde is used to distinguish it from the standard scattering amplitude $\mathcal{A}$. Furthermore, $\tilde{\mathcal{A}}$ contains propagators with a retarded $i \epsilon$ prescription. 
In other words: if $p$ is taken on shell, $\tilde A$ is a standard amplitude, up to the $i\epsilon$ prescription, for appropriate choices of sign of the energies $q_1^0$ and $q_2^0$.

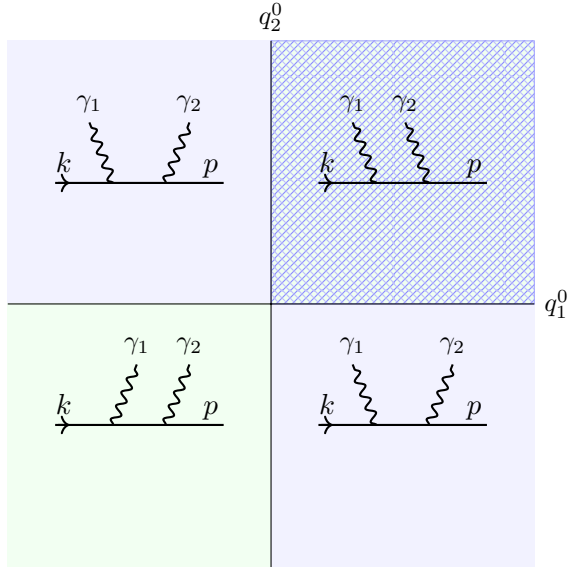
\begin{figure}
    \centering
    \begin{tikzpicture}[scale=0.70,transform shape
  axis/.style={thin},
  scalar/.style={thick},
  photonbase/.style={thick, decorate, decoration={snake, amplitude=1.6pt, segment length=5.5pt},-},
  photonout/.style={photonbase, -Latex,-},
  photonin/.style={photonbase, Latex-,-},
  blob/.style={circle, fill=allOrderBlue!25, draw=black, line width=1.1pt, minimum size=0.0pt, inner sep=0pt,fill opacity=1,
  draw opacity=1},
  lab/.style={font=\small}
]

\fill[blue!5] (-5,0) rectangle (0,5);      
\fill[green!5] (-5,-5) rectangle (0,0);     
\fill[blue!5] (0,-5) rectangle (5,0);      

\fill[green!6] (0,0) rectangle (5,5);       
\begin{scope}
  \clip (0,0) rectangle (5,5);
  \path[pattern=north east lines, pattern color=blue!45] (0,0) rectangle (5,5);
  \path[pattern=north west lines, pattern color=blue!25] (0,0) rectangle (5,5);
\end{scope}
\draw[blue!35, line width=0.4pt] (0,0) rectangle (5,5);

\draw (-5,0) -- (5,0) node[right, lab] {$q_1^{0}$};
\draw (0,-5) -- (0,5) node[above, lab] {$q_2^{0}$};
\fill (0,0) circle[radius=0.6pt];

\newcommand{\QuadDiagram}[4]{%
  \begin{scope}[shift={(#1,#2)}]
    \coordinate (sL) at (-1.6,0);
    \coordinate (v1) at (-0.5,0);
    \coordinate (v2) at (0.5,0);
    \coordinate (sR) at (1.6,0);

   \draw[->, thick]
  ($(sL)!0.05!(v1)$) -- ($(sL)!0.25!(v1)$)
  node[midway, above] {$k$} ;

\path (v2) -- (sR) node[ pos=0.78,above=-1.7pt] {$p$};

    \coordinate (g1out) at (-0.05,1.15);
    \coordinate (g1in)  at (-0.95,1.15);
    \coordinate (g2out) at ( 0.95,1.15);
    \coordinate (g2in)  at ( 0.05,1.15);

    \if#3O
      \draw[photonout] (v1) -- (g1out) node[lab, above] {$\gamma_1$};
    \else
      \draw[photonin]  (v1) -- (g1in)  node[lab, above] {$\gamma_1$};
    \fi

    \if#4O
      \draw[photonout] (v2) -- (g2out) node[lab, above] {$\gamma_2$};
    \else
      \draw[photonin]  (v2) -- (g2in)  node[lab, above] {$\gamma_2$};
    \fi

    \draw[scalar] (sL) -- (v1) -- (v2) -- (sR);
    \node[blob] at (v1) {};
    \node[blob] at (v2) {};
  \end{scope}%
}

\QuadDiagram{ 2.5}{ 2.3}{I}{I} 
\QuadDiagram{-2.5}{ 2.3}{I}{O} 
\QuadDiagram{-2.5}{-2.3}{O}{O} 
\QuadDiagram{ 2.5}{-2.3}{I}{O} 

\end{tikzpicture}
\caption{Schematic representation of $\tilde{\mathcal{A}}(k,q_1,q_2,p)$ in the four $q_i^0$ quadrants, with time flowing from left to right. (Symmetrization of contributions, as in (\ref{eq:OffShellAmplitude}), is implicit.) The diagrams in the blue quadrants vanish when $p$ is taken to be incoming, while that in the green quadrant vanishes when $p$ is outgoing. The diagram in the shaded quadrant vanishes for both configurations.\label{fig:quadrants}}
\end{figure}

We are now ready to discuss how to pass from $R^{(2)}(-k,p)$ to the Bogoliubov coefficients at $\mathcal{O}(g^2)$ order, by placing $\pm p$ on-shell. Prior to this, it is convenient to split the integration measure into four quadrants in the $q$-energy-planes as follows:
\[
\hd^4 q_1 \, \hd^4 q_2 = \hd^4 q_1 \, \hd^4 q_2 \left( \theta(q_1^0)\theta(q_2^0) + \theta(q_1^0)\theta(-q_2^0) + \theta(-q_1^0)\theta(q_2^0)+ \theta(-q_1^0)\theta(-q_2^0)\right)\,.
\]
This, in turn, leads us to the following natural definition of the sector-wise contributions to the response function (see Fig.~\ref{fig:quadrants}):
\begin{align}\label{eq:CoherentResponse2}
    R_{\pm\pm}^{(2)}(-k,p)= \frac{1}{2}\int  \hd^4 q_2\,\hd^4 q_1 \,\theta(\pm q_1^0)\theta(\pm q_2^0) \, \hat{\delta}^4(k+q_1+q_2-p)\,\tilde{\Gamma}(q_2)\tilde{\Gamma}(q_1)\, i\tilde{\mathcal{A}}(k , q_1 , q_2 ,p)\,. 
\end{align}
We now show that, upon placing $\pm p$ on-shell, kinematics restrict the contributions to the Bogoliubov coefficients to specific sectors only.

We start by considering $\alpha^{(2)}(p,k)$, the $\mathcal{O}(g^2)$ contribution to the Bogoliubov $\alpha(p,k)$, obtained by placing $p$ on-shell in $R^{(2)}(-k,p)$, which ultimately amounts to the replacement $\tilde{\mathcal{A}}\rightarrow \mathcal{A}$. First we focus on the $++$ and $--$ sectors, which correspond to the amplitudes $ \mathcal{A}(k,q_1,q_2\rightarrow p)$ and $\mathcal{A}(k\rightarrow p,q_1,q_2)$, respectively. For both sectors, the on-shell condition for $p$ cannot be satisfied for arbitrary momenta, resulting in a vanishing contribution. On the other hand, $+-$ and $-+$ sectors---contributing equally by symmetry---capture the surviving contribution to $\alpha^{(2)}(p,k)$, yielding
\[\label{eq:alph2_from_A}
\alpha^{(2)}(p,k)-\delta_\Phi(k-p)
&= \left[R^{(2)}_{+-}(-k,p) + R^{(2)}_{-+}(-k,p)\right]_{p^2=m^2} \\
&=   \int \dd \Phi(q_1) \, \dd \Phi(q_2)\, \gamma(q_1)\gamma^*(q_2)
\, i \mathcal{A}(k,q_1\rightarrow p,q_2
) \hdelta^4(k+q_1 - p - q_2)\,,
\]
where we have changed variables $q_2 \rightarrow -q_2$ and used $\gamma(-q_2) = \gamma^*(q_2)$.

We now turn to $\beta^{(2)}(p,k)$, the $\mathcal{O}(g^2)$ contribution to the Bogoliubov $\beta(p,k)$, which arises by placing $p$ on-shell in $R^{(2)}(-k,-p)$. In this case, $p$ is incoming and the only surviving contribution comes from $R^{(2)}_{--}(-k,-p)$, in which both $q_1$ and $q_2$ are outgoing. Therefore
\[\label{eq:beta2_from_A}
-\beta^{(2)*}(p,k)
&= \left[R^{(2)}_{--}(-k,-p)\right]_{p^2=m^2} \\
&= -\frac{1}{2} \int \dd \Phi(q_1) \, \dd \Phi(q_2) \gamma^*(q_1)\gamma^*(q_2) i \mathcal{A}(k,p\rightarrow q_1,q_2
) \hdelta^4( k + p-q_1 - q_2).
\]
Comparing (\ref{eq:alph2_from_A}) and (\ref{eq:beta2_from_A}), a few remarks are in order. First, it is straightforward to extend the above calculation to $\mathcal{O}(g^n)$, for which see Appendix~\ref{AppendixA}. Second, the above calculation admits a straightforward generalization to backgrounds with non-zero spin --  this effectively amounts to replacing\footnote{We note that ${\tilde\Gamma}_\eta(-q) = - {\tilde \Gamma}^*_\eta(q)$.} $\Gamma$ with $\Gamma_{\eta}$, and the vacuum off-shell current $\tilde{\mathcal{A}}(k,q_1,q_2,p)$ with the appropriate object $\tilde{\mathcal{A}}(k,p,q^{\eta_1}_1,q^{\eta_2}_2)$.

What is worth highlighting is that while the crossing relation between $\alpha(p,k)$ and $\beta^*(p,k)$ involves only a single leg (i.e. $p\rightarrow -p$), from \eqref{eq:alph2_from_A} and \eqref{eq:beta2_from_A} we find that once we expand these quantities in terms of vacuum amplitudes, an implicit two-leg crossing -- namely $\mathcal{A}(k,q_1\rightarrow p,q_2)\rightarrow \mathcal{A}(k,p\rightarrow q_1,q_2)$ -- is at work beneath the momentum integrals. The emergence of this two-leg crossing suggests a direct connection between crossing in background-field QFT and the standard crossing property of vacuum scattering amplitudes discussed in e.g.~\cite{Caron-Huot:2023ikn}. We will now discuss the details of how this works, using scalar QED to illustrate.

\subsection{Connection to standard on-shell crossing}
\label{sec:Standard_Crossing}

The analogue of $\tilde{\mathcal{A}}$ in scalar QED is $\tilde{\mathcal{A}}(k,p,q^{\eta_1}_1,q^{\eta_2}_2)$, with $\eta_{i}$ denoting the photon helicity. 
For the evaluation of $\alpha^{(2)}$, therefore, the appropriate object is the scalar QED version of \eqref{eq:alph2_from_A}, namely the leading order Compton amplitude, with contributing diagrams
\[
\begin{tikzpicture}[scale=0.8, baseline={([yshift=-\the\dimexpr\fontdimen22\textfont2\relax] current bounding box.center)}]
	\usetikzlibrary{decorations.pathmorphing}
	\tikzset{snake it/.style={decorate, decoration=snake}}
	\path [draw=black, thick] (3.5,-0.5) -- (6.5,-0.50);
	\path [draw=black, snake it, thick] (3.5,-1.5) -- (5.0,-0.5);
    \path [draw=black, snake it, thick] (6.5,-1.5) -- (5.0,-0.5);
    \filldraw[fill=allOrderBlue!25,very thick] (5,-0.5) circle (4.0pt);
\end{tikzpicture}
=
\begin{tikzpicture}[scale=0.8, baseline={([yshift=-\the\dimexpr\fontdimen22\textfont2\relax] current bounding box.center)}]
	\usetikzlibrary{decorations.pathmorphing}
	\tikzset{snake it/.style={decorate, decoration=snake}}
	\path [draw=black, thick] (3.5,-0.5) -- (6.5,-0.50);
	\path [draw=black, snake it, thick] (3.5,-1.5) -- (4.5,-0.5);
    \path [draw=black, snake it, thick] (6.5,-1.5) -- (5.5,-0.5);
\end{tikzpicture}
\;
+
\begin{tikzpicture}[scale=0.8, baseline={([yshift=-\the\dimexpr\fontdimen22\textfont2\relax] current bounding box.center)}]
	\usetikzlibrary{decorations.pathmorphing}
	\tikzset{snake it/.style={decorate, decoration=snake}}
	\path [draw=black, thick] (3.5,-0.5) -- (6.5,-0.50);
	\path [draw=black, snake it, thick] (3.5,-1.5) -- (5.75,-0.5);
    \path [draw=black, snake it, thick] (6.5,-1.5) -- (4.25,-0.5);
\end{tikzpicture}
\;
+
\begin{tikzpicture}[scale=0.8, baseline={([yshift=-\the\dimexpr\fontdimen22\textfont2\relax] current bounding box.center)}]
	\usetikzlibrary{decorations.pathmorphing}
	\tikzset{snake it/.style={decorate, decoration=snake}}
	\path [draw=black, thick] (3.5,-0.5) -- (6.5,-0.50);
	\path [draw=black, snake it, thick] (3.5,-1.5) -- (5.0,-0.5);
    \path [draw=black, snake it, thick] (6.5,-1.5) -- (5.0,-0.5);
\end{tikzpicture}
\; \;.
\]
These correspond to $s,u$-channels and the seagull, such that
\begin{align}
    {\cal A}(k,q_1^{\eta_1}\to p,q_2^{\eta_2}) = -2g^2 
    \left[
    \frac{N_s}{s-m^2+i\epsilon} +
    \frac{N_u}{u-m^2+i\epsilon} - 
    \varepsilon_{\eta_1}(q_1)\cdot \varepsilon^{\dagger}_{\eta_2}(q_2)
    \right],
\end{align}
in which the numerators are
\begin{align}
    N_s = 2\varepsilon_{\eta_1}(q_1)\cdot k\,\,\varepsilon^{\dagger}_{\eta_2}(q_2)\cdot (k+q_1), \qquad
    N_u = 2\varepsilon^{\dagger}_{\eta_2}(q_2)\cdot k\,\,\varepsilon_{\eta_1}(q_1)\cdot(k-q_2),
\end{align}
and the Mandelstam invariants are
\begin{align}
   s = (k+q_1)^2,\qquad u = (k-q_2)^2, \qquad \text{and}\qquad  t = (k-p)^2.
\end{align}
We now consider the two leg-crossing $\mathcal{A}(k,q^{\eta_1}_1\rightarrow p,q^{\eta_2}_2)\rightarrow \mathcal{A}(k,p\rightarrow q^{\eta_1}_1,q^{\eta_2}_2)$ that gives rise to, in the background QFT picture, the single leg crossing $p\rightarrow -p$ relating $\alpha^{(2)}$ and $\beta^{(2)}$. In order to cross the particles $p$ and $q_1$ in the amplitude $ {\cal A}(k,q_1^{\eta_1}\to p,q_2^{\eta_2})$ above we follow the prescription of~\cite{Caron-Huot:2023ikn,Mizera:2023tfe}, first deforming the momenta in light-cone coordinates as follows:
\begin{align}
    p^\mu(z) = \left(z p^+,\tfrac{1}{z}p^- , p^\perp\right),
    \qquad
    {q_1}^\mu(z) = \left(z q_1^+,\tfrac{1}{z} q_1^- , q_1^\perp\right),
\end{align}
where $p^\pm = p^0\pm p^3$ and $p^\perp=(p^1,p^2)$. This choice keeps the $u$-channel fixed, keeps the particles on-shell ($p^2(z) = m^2$, $q_1^2(z) = 0$) and conserves momentum $k+q_1(z)=p(z)+q_2$.

Crossing through a path along a large arc $|z|\gg 1$ accounts for flipping the energies, similar to the previous section, and the resulting amplitude is that of pair annihilation. Diagrammatically, this is represented by
\[
\Bigg[
\quad
\begin{tikzpicture}[scale=0.8, baseline={([yshift=-\the\dimexpr\fontdimen22\textfont2\relax] current bounding box.center)}]
	\usetikzlibrary{decorations.pathmorphing}
	\tikzset{snake it/.style={decorate, decoration=snake}}
	\path [draw=black, thick] (3.5,-0.5) -- (6.5,-0.50);
	\path [draw=black, snake it, thick] (3.5,-1.5) -- (5.0,-0.5);
    \path [draw=black, snake it, thick] (6.5,-1.5) -- (5.0,-0.5);
    \filldraw[fill=allOrderBlue!25,very thick] (5,-0.5) circle (4.0pt);
\end{tikzpicture}
\quad
\Bigg]_{\curvearrowleft}
\;
\qquad
\Rightarrow
\qquad
\;
\begin{tikzpicture}[scale=0.8, baseline={([yshift=-\the\dimexpr\fontdimen22\textfont2\relax] current bounding box.center)}]
	\usetikzlibrary{decorations.pathmorphing}
	\tikzset{snake it/.style={decorate, decoration=snake}}
	\path [draw=black, snake it,thick] (5.0,-0.5) -- (6.5,-0.5);
    \path [draw=black, snake it,thick] (5.0,-0.5) -- (6.5,-1.5);
	\path [draw=black, thick] (3.5,-0.5) -- (5.0,-0.5);
    \path [draw=black, thick] (3.5,-1.5) -- (5.0,-0.5);
    \filldraw[fill=allOrderBlue!25,very thick] (5,-0.5) circle (4.0pt);
\end{tikzpicture}
\;
\]
The crossing of the photon leg induces a Wigner rotation~\cite{Trueman:1964zzb,Hara:1970gc,Hara:1964zza}, but this is merely a phase. A basis for the polarization vectors can be chosen such that they cross as $\varepsilon^\mu_{\eta_1}(-q_1) = -(\varepsilon^\mu_{\eta_1}(q_1))^*$, see Section 2 of \cite{Hara:1964zza} ($\varepsilon_{\eta_2}(q_2)$ remains unchanged). The poles of the amplitude in $s$ and $t$ deform to
\[
[s(z)-m^2] &= (p(z) + q_2)^2-m^2 \simeq  zp^+ q_2^- ,
\\
[t(z)-m^2] &= (k-p(z))^2-m^2 \simeq -z p^+ k^-,
\]
in lightcone coordinates. In the Lorentz frame, $p^\pm = q_1^\pm$ also implies that $k^\pm = q_2^\pm$ due to momentum conservation. It is easy to see that there is a path in this crossing since $s(z) -m^2\simeq  zp^+ q_2^- = zp^+k^-$ while $t(z)-m^2\simeq -z p^+ k^- =-zp^+q_2$, and the analytical continuation can be performed along a large arc $|z|\gg 1$. However, this path from $s(z)$ to $t(z)$ ends on the `wrong' side of the real axis (see~\cite{Mizera:2023tfe}).

After this crossing, the amplitude is 
\[
    \label{eq:Crossing_Flat_Amps}
    \left[{\cal A}(k,q_1^{\eta_1}\to p,q_2^{\eta_2})\right]_{\curvearrowleft} 
    &= 2g^2 
    \left[
    \frac{N^{'}_s}{t-m^2-i\epsilon} +
    \frac{N^{'}_u}{u-m^2+i\epsilon} 
    - \varepsilon^\dagger_{\eta_1}(q_1)\cdot \varepsilon^{\dagger}_{\eta_2}(q_2)
    \right], \\
    &=\left[-{\cal A}(k,p \to q_1^{\eta_1},q_2^{\eta_2})\right] + \text{cut}
\]
where the $N^{'}_{s,u}$ numerators accounts for $\varepsilon_{\eta_1}(q_1) \to \varepsilon^*_{\eta_1}(-q_1) = -\varepsilon^*_{\eta_1}(q_1)$ and the momenta deformation. The resulting amplitude is that of pair annihilation, which contributes to the Bogoliubov coefficient $\beta^{(2)*}(p,k)$ in \eqref{eq:beta2_from_A}. Note that the crossing introduces an additional cut to the amplitude that deforms the $i\epsilon$ prescription. (The minus sign is consistent with the parity transformation of ${\tilde \Gamma}_\eta$.) Indeed, this is precisely the relationship we expect considering the general crossing equation between $\alpha^*(p,k)-  \delta_\Phi(p-k)$ and $-\beta^*(p,k)$.

We conclude that equation \eqref{eq:BogoliubovCrossingEquation}
(see also Eq.\! 3.4 in~\cite{Caron-Huot:2023ikn})
arises from crossing relations between perturbative flat-space amplitudes as in \eqref{eq:Crossing_Flat_Amps} when the background may be described as a coherent state. This can be extended to higher orders in the coupling by including contributions of larger numbers of coherent-state particles.

\section{Conclusion}
\label{sec:Conclusion}
We have revisited quantum field theory in background fields from an explicitly on-shell perspective. Focusing on probe fields interacting quadratically with a fixed background, we have shown that the familiar Bogoliubov coefficients admit a natural interpretation as generalised amplitudes with in-in boundary conditions. In this language, particle scattering (encoded in the standard amplitudes) and mode mixing (encoded in the Bogoliubov coefficients), in a fixed background, are but different facets of a Gaussian $S$-matrix structure.

We discussed how all physical observables in the probe sector are controlled by a small set of primitive processes: one-to-one scattering and pair creation/annihilation. These processes assemble into the full background $S$-matrix, which acts as a squeezing operator on the in-vacuum. This structure is well known from the operator approach to background QFT, but recasting it in amplitudes language clarifies several conceptual points. In particular, the Bogoliubov coefficients $\alpha$ and $\beta$ are seen to be on-shell objects, which in turn are related to the aforementioned amplitudes via formally exact functional expressions.

Building on this, we examined how crossing symmetry manifests in this setting. We identified the underlying principles that endow on-shell quantities in QFT in fixed backgrounds with well-defined analytic properties, even while these objects depend on arbitrary background functions. We showed that the Bogoliubov coefficients can be obtained from the LSZ reduction of retarded correlators, and that it is precisely the causal structure of these retarded products that gives rise to analyticity in their truncated Fourier transforms, which we call the response function. This, in turn, leads to crossing relations between $\alpha$ and $\beta$. For backgrounds of finite support in time (at least), the primitive 1-to-1 amplitude and the pair creation amplitude are likewise related by a crossing relation, again following from the analyticity of truncated Fourier transforms of the time-ordered correlator.

These results are non-perturbative, in the sense that the coupling between the probe and the background is treated exactly. Complementary insights were nevertheless achieved by examining our various on-shell objects when the interaction with the background is handled perturbatively. This approach makes it explicit that the distinction between scattering amplitudes and Bogoliubov coefficients is naturally tied to the choice of propagator -- Feynman for the former and retarded for the latter. The non-trivial relations between Bogoliubov coefficients on the one hand and primitive amplitudes on the other then follow directly from this distinction, using the simple relation connecting the two propagators.

For background fields generated by a coherent state, we showed how the crossing relations for Bogoliubov coefficients are connected to crossing in vacuum amplitudes. In particular, we clarified how a seemingly single-leg crossing in the background QFT context emerges from a two-leg crossing of vacuum amplitudes, using scalar QED as an example.

It is satisfying to see that vacuum crossing relations underlie and give rise to the corresponding crossing behaviour in background QFT. We expect this crossing relation to persist at higher loop orders, although this has only been rigorously shown for theories with a mass gap~\cite{Bros:1964iho,Bros:1965kbd, Bros:1972jh,BROS_1986325}: the presence of infrared divergences in scattering amplitudes with massless particles appears to obstruct the necessary analytic continuation.

Indeed, the addition of loop corrections is one possible direction for future research. Others include the incorporation of backreaction effects, allowing quantum dynamics of the background, and investigating crossing and causality of amplitudes in systems with boundaries/horizons or non-trivial topology.

\acknowledgments

We thank Matteo Sergola for helpful comments. AI, DOC, and KR are supported by the STFC Consolidated Grant ST/X000494/1 ``Particle Theory at the Higgs Centre".
RA is supported by UKRI under the UK government’s Horizon Europe Marie Sklodowska Curie funding guarantee grant EP/Z000947/1.
DOC is supported by the European Research Council under Advanced Investigator grant ERC–AdG–101200505.

\appendix

\section{The $n$-point response function}
\label{AppendixA}

Here we outline the details of the $\mathcal{O}(g^n)$ contribution to the Bogoliubov coefficients and the crossing relations underlying them by extending the calculations in Section~\ref{sec:coherent_state_X}.

We start with the straightforward extension of \eqref{eq:R_g2_scalar} to higher multiplicity amplitudes, and follow the same steps that lead to \eqref{eq:CoherentResponse2}, which yields
\[
R^{(n)}_{\pm,\cdots,\pm} = \frac{1}{n!}\left(\prod_{i=1}^n \int \hd^4 q_{i} \,\theta(\pm q_i^0)\tilde{\Gamma}(q_i) \right)i \tilde{\mathcal{A}}(k,q_1,\cdots,q_n,p)\hdelta^4\left(k-p+\sum_{i=1}^n q_i\right)\,.
\]
Applying the LSZ reduction and summing over the $2^N$ choices for the signs of $q_i^0$ we obtain the following expressions for the Bogoliubov coefficients:
\[\label{eq:CoherentAlphaN}
\alpha^{(n)}(p,k) = \frac{1}{n!} \sum_{m=0}^n \binom{n}{m} \prod_{i=1}^n \int \hd \Phi(q_i) (\gamma(q_1) \cdots \gamma(q_m))(\gamma^*(q_{m+1}) \cdots \gamma^*(q_n))\\
\times i\mathcal{A}(k,q_1 \cdots q_m \rightarrow p, q_{m+1}, \cdots q_n) \hdelta^4\left(k-p+\sum_{i=1}^n q_i\right)
\]
where we have summed over all ways of picking $m$ incoming particles from the set $\{q_1,\cdots ,q_n\}$. Note that this sum includes terms that vanish due to the momentum conservation and on-shell constraints. Similarly, we get the following expression for $\beta^{(n)*}(p,k)$
\[\label{eq:CoherentBetaN}
\beta^{(n)*}(p,k) = \frac{1}{n!} \sum_{m=0}^n \binom{n}{m} \prod_{i=1}^n \int \hd \Phi(q_i) (\gamma(q_1) \cdots \gamma(q_m))(\gamma^*(q_{m+1}) \cdots \gamma^*(q_n))\\
\times i\mathcal{A}(k,p,q_1 \cdots q_m \rightarrow  q_{m+1}, \cdots q_n) \hdelta^4\left(k-p+\sum_{i=1}^n q_i\right).
\]
We can now verify the prediction of \eqref{eq:CoherentBogoliubov} by checking that the coherent state expansions reproduce the same combinatoric factors in \eqref{eq:CoherentAlphaN} and \eqref{eq:CoherentBetaN}. Explicitly, we expand \eqref{eq:CoherentBogoliubov} to order $g^n$ to find
\[
\alpha^{(n)}(p,k) = \sum_{m=0}^n \frac{1}{m! (n-m)!} \int \hd \Phi(q_1, \cdots, q_n) (\gamma(q_1) \cdots \gamma(q_m))(\gamma^*(q_{m+1}) \cdots \gamma^*(q_n))\\
\times \langle q_1, \cdots q_m| S^\dagger a(p) S a^\dagger(k) |q_{m+1} \cdots q_n \rangle^{(n)},
\]
and
\[
\beta^{(n)*}(p,k) = \sum_{m=0}^n \frac{1}{m! (n-m)!} \int \hd \Phi(q_1, \cdots, q_n) (\gamma(q_1) \cdots \gamma(q_m))(\gamma^*(q_{m+1}) \cdots \gamma^*(q_n)) \\
\times   \langle q_1, \cdots q_m|a(k)S^\dagger a(p) S |q_{m+1} \cdots q_n \rangle^{*,(n)}.
\]
The coefficients $1/(m! (n-m)!)$ are obtained from the coherent state exponential. It is easy to check that this coefficient agrees with the expressions \eqref{eq:CoherentAlphaN} and \eqref{eq:CoherentBetaN} by noting that
\[
 \frac{1}{n!} \binom{n}{m} = \frac{1}{m! (n-m)!}.
\]
It remains to verify that
\[
 \langle q_1, \cdots q_m| S^\dagger a(p) S a^\dagger(k) |q_{m+1} \cdots q_n \rangle^{(n)} = i\mathcal{A}(k,q_1 \cdots q_m \rightarrow p, q_{m+1}, \cdots q_n) \hdelta^4\left(k-p+\sum_{i=1}^n q_i\right)\\
  \langle q_1, \cdots q_m|a(k)S^\dagger a(p) S |q_{m+1} \cdots q_n \rangle^{*,(n)} =  i\mathcal{A}(k,p,q_1 \cdots q_m \rightarrow  q_{m+1}, \cdots q_n) \hdelta^4\left(k-p+\sum_{i=1}^n q_i\right)
\]
To do this, we note that the left-hand side computes the amplitudes $i\mathcal{A}(k,q_1 \cdots q_m \rightarrow p, q_{m+1}, \cdots q_n)$ and $i\mathcal{A}(k,p,q_1 \cdots q_m \rightarrow  q_{m+1}, \cdots q_n)$ with a retarded $i\epsilon$ prescription. Which is precisely what is obtained from the response function calculation above

\bibliographystyle{JHEP}
\bibliography{Crossing-causality}

\end{document}